\documentclass[twocolumn]{aastex62}

\usepackage{mathrsfs}

\makeatletter
\def\cleardoublepage{\clearpage\if@twoside \ifodd\c@page\else
\hbox{}
\thispagestyle{empty}
\newpage
\if@twocolumn\hbox{}\newpage\fi\fi\fi}
\makeatother 
\newcommand{\BAYMAX}{{\tt BAYMAX}}

\shortauthors{Foord et al.}

\usepackage{relsize}

\usepackage{relsize}
\makeatletter
\providecommand{\ion}[2]{#1$\;$\textsmaller{\@Roman{#2}}}
\makeatother

\usepackage{amsmath,amssymb} 
\usepackage{natbib} 

\newcommand{\beq}{
\begin{equation}
}
\newcommand{\eeq}{
\end{equation}
}

\newcommand{\beqa}{
\begin{eqnarray}
}
\newcommand{\eeqa}{
\end{eqnarray}
}

\begin{document}

\title{AGN Triality of Triple Mergers: Multi-wavelength Classifications}

\author[0000-0002-1616-1701]{Adi Foord}
\affil{Department of Astronomy and Astrophysics, University of Michigan, Ann Arbor, MI 48109}
\affil{Kavli Institute of Particle Astrophysics and Cosmology, Stanford University, Stanford, CA 94305}

\author[0000-0002-1146-0198]{Kayhan G\"{u}ltekin}
\affil{Department of Astronomy and Astrophysics, University of Michigan, Ann Arbor, MI 48109}

\author{Jessie C. Runnoe}
\affil{Department of Physics and Astronomy, Vanderbilt University, Nashville, TN 37235}

\author{Michael J. Koss}
\affil{Eureka Scientific Inc, Oakland, CA, 94602}



\begin{abstract}
We present results from a multi-wavelength analysis searching for multiple AGN systems in nearby ($z<0.077$) triple galaxy mergers. Combining archival \emph{Chandra}, SDSS, \emph{WISE}, and VLA observations, we quantify the rate of nearby triple AGN, as well as investigate possible connections between SMBH accretion and merger environments. Analyzing the multi-wavelength observations of 7 triple galaxy mergers, we find that 1 triple merger has a single AGN (NGC 3341); we discover, for the first time, 4 likely dual AGN (SDSS J1027+1749, SDSS J1631+2352, SDSS J1708+2153, and SDSS J2356$-$1016); we confirm one triple AGN system, SDSS J0849+1114; and 1 triple merger in our sample remains ambiguous (SDSS J0858+1822). Analyzing the \emph{WISE} data, we find a trend of increasing $N_{H}$ (associated with the primary AGN) as a function of increasing $W1$--$W2$ color, reflecting that the motions of gas and dust are coupled in merging environments, where large amount of both can be funneled into the active central region during mergers. Additionally, we find that the one triple AGN system in our sample has the highest levels of $N_{H}$ and $W1$--$W2$ color, while the dual AGN candidates all have lower levels; these results are consistent with theoretical merger simulations that suggest higher levels of nuclear gas are more likely to activate AGN in mergers. 
\end{abstract}

\section{Introduction}
\label{chap6:intro}
Systems with multiple active galactic nuclei (AGN) are theorized to be relatively common in the nearby Universe \citep{Kelley2017a, Ryu2018}, and are expected as a result of hierarchical galaxy formation (e.g., \citealt{WhiteandReese1978, MilosavljevicMerritt2003, Barnes1991, DiMatteo2008, Angles2017}). Studying the properties of multiple AGN systems, in a systematic manner, is important given the unknown magnitude that mergers play in SMBH growth and evolution. 
\par There is strong reason to believe that interplay between host galaxies and their respective SMBHs during mergers exist. Galaxy-mergers are thought to be a key process behind the various empirical SMBH-galaxy scaling relations, such as the relation between the supermassive black hole mass and host-galaxy bulge velocity dispersion ($M$--$\sigma$ relation) and luminosity ($M$--$L$ relation; \citealt{Magorrian1998, Ferrarese2000, Tremaine2002, Gultekin2009b, McConnell2013}). These relations likely arise due to a combination of triggered feedback processes from the AGN \citep{Hopkins2006}, triggered star formation \citep{Sorbral2015}, or possibly repeated mergers \citep{JahnkeandMaccio2011}. Yet, the connection between mergers and SMBH activity remains poorly understood; various studies have found conflicting results regarding whether mergers are responsible for, or even correlated with, SMBH activity \citep{Urrutia2008, Koss2010, Ellison2011, Treister2012, Schawinski2012, Ellison2013, Villforth2014, Satyapal2014, Glikman2015, Fan2016, Weston2017, Goulding2017, Onoue2018, Barrows2018, Koss2018}. Past measurements were likely complicated by (1) AGN variability (see, e.g., \citealt{Goulding2017, Capelo2017}, (2) the difficulty in measuring higher-redshift mergers \cite{Hung2014}, and (3) the obscuration and merger-stage dependency of AGN activity \cite{Veilleux2002, Koss2012, Kocevski2015}. Thus, one of the best ways to analyze the possible ties between merger environments and SMBH activity is to study systems with unique observational flags of merger-driven SMBH growth --- such as multiple AGN. Furthermore, studying multiple AGN systems in X-rays, which are less affected by obscuration than optical diagnostics \cite{Koss2017} and represent a larger fraction of the AGN population than radio-loud AGN, will result in the most complete study. 
\par In our previous study searching for multiple AGN systems (Foord et al. 2020b, \emph{submitted}), we analyzed the X-ray observations of a sample of 7 nearby ($z<0.077$) triple merging galaxies. Due to possibly high levels of dust in merging environments (e.g., \citealt{Hopkins2005, Kocevski2015, Koss2016, Ricci2017, Blecha2018, DeRosa2018, DeRosa2019}), X-rays are a powerful diagnostic to finding AGN whose optical emission may be diluted (e.g., \citealt{Koss2011}). One of the larger goals of the study was to find new nearby triple AGN, which are theorized to play important roles in the coalescence of SMBHs and the stochastic gravitational wave background (GWB). Simulations show that interactions from a tertiary SMBH can assist two interacting SMBHs to merge within a Hubble time (otherwise they may be stalled, see \citealt{Milosavljevic2003}), reducing the merger time by over a factor of 10 \citep{Blaes2002}. Since the SMBH binary population driving the detectable GWB signal is expected to be relatively nearby ($z<0.3$; \citealt{Kelley2017a}), searching for nearby triple AGN was of particular interest. 
\par In Foord et al. (2020b), we analyzed the \emph{Chandra} observations of the 7 triple merging galaxies with the tool \BAYMAX{} (Bayesian AnalYsis of Multiple AGN in X-rays), in order to find faint multiple AGN systems that may otherwise go undetected. \BAYMAX{} uses a Bayesian framework (namely, Bayes factor) to quantify whether a given \emph{Chandra} observation is better described by one or multiple point sources (see \citealt{Foord2019, Foord2020a} for explicit details). By comparing the Bayes factors between single, dual, and triple X-ray point source models,  we found 1 triple merger composed of a single X-ray point source; 5 triple mergers likely composed of two X-ray point sources; and one system composed of three X-ray point sources. By fitting the individual X-ray spectra of each point source, we analyzed the $2-7$ keV luminosities and found that 4 out of 6 dual X-ray point source systems have primary and secondary point sources with bright X-ray luminosities ($L_{\mathrm{2-7 \ keV}}>10^{40}$erg s$^{-1}$). Classifying X-ray point sources with $L_{{\mathrm{2-7 \ keV}}} > 10^{41}$ erg s$^{-1}$ as bona fide AGN, we further concluded that 6 of the X-ray point sources detected were AGN. However, a multi-wavelength analysis is required for a better understanding of the true duality, or triality, of all these systems. 
\par In this paper, we combine the X-ray results with archival Sloan Digital Data Survey (SDSS), \emph{Wide-field Infrared Survey Explorer} (\emph{WISE}), and Very Large Array (VLA) observations, in order to better classify each X-ray point source, learn more about the preferential environments of each multiple AGN candidate, and compare AGN classifications between X-ray and optical AGN diagnostics. In particular, because our X-ray classifications (as described in Section~\ref{chap6:origin}) of an AGN may differ from other multi-wavelength diagnostics, comparing results from multiple methods can lend insight into physical parameters such as star formation rates, obscuration, and Eddington fractions (see Section~\ref{chap6:discussion}). 
\par The remainder of the paper is organized into 4 sections. In Section~\ref{chap6:sample} we review the sample and the existing multi-wavelength coverage. In Section~\ref{chap6:origin} we classify the nature of accretion for each X-ray point source, using archival IR observations to estimate the levels of X-ray emission expected from X-ray binaries intrinsic to the host galaxies. In Section~\ref{chap6:discussion} we compare environmental properties between the single, dual, and triple AGN. Lastly, we summarize our findings in Section~\ref{chap6:conclusions}. Throughout the paper we assume a $\Lambda$CDM universe, where $H_{0}=69.6$, $\Omega_{M}=0.286$, and $\Omega_{\Lambda}=0.714$.
%
\section{Sample}
\label{chap6:sample}
Details regarding our sample selection, as well as specifics of the \emph{Chandra} programs for each observation, are thoroughly outlined in Foord et al. (2020b). However we briefly review the selection criteria and summarize the X-ray properties below.
\subsection{Sample Selection}
We first cross-match all nearby ($z<0.3$) AGN from the AllWISE AGN catalog \citep{Secrest2015} with the SDSS Data Release 16 (SDSS DR16) catalog \citep{SDSSDR16}. Using the SDSS DR16 data, we visually identify any system that is part of a triple galaxy merger, enforcing the criteria that (i) a photometric or spectroscopic redshift measurement is available for each galaxy in a triple merger system and (ii) that the respective redshifts of each galaxy in a triple merger system are consistent with one another at the 3$\sigma$ confidence level. Of the 12 systems that meet these criteria, 4 have existing, on-axis, archival \emph{Chandra} observations (SDSS J1708+2153, SDSS J2356$-$1015, SDSS J1631+2252, SDSS J0849+1114), and we add 3 triple galaxy mergers from the literature with archival \emph{Chandra} and SDSS DR16 observations that meet our redshift criteria: NGC 3341 \citep{Bianchi2013}, SDSS J0858+1822 \citep{Liu2011} and SDSS J1027+1749 \citep{Liu2011b}. 
\par Given our selection process, the separations between all galactic nuclei range from 3$-$15 kpc. All of the systems in our sample have archival \emph{Chandra}, \emph{WISE}, and SDSS DR16 observations, while three (SDSS J0849+1114, SDSS J0858+1822, SDSS J1027+1749) have additional multiband HST/WFC3 (F506W and F5336W) imaging (PI: Liu, Proposal ID: 13112). In Table~\ref{chap6:tabTripleGalInfo} we list the galaxy sample properties.

\subsection{X-ray Properties}
Each of the \emph{Chandra} observations was previously analyzed with \BAYMAX~(Foord et al. 2020b). It was found that 1 triple merger system favored the single point source model (SDSS J0858+1822, although, given the low 2$-$7 keV luminosity and relatively soft spectrum, it was classified as non-AGN emission); 5 triple merger systems favored the dual point source model (SDSS J1027+1749, NGC 3341, SDSS J1631+2352, SDSS J1708+2153, and SDSS J2356$-$1016); and one triple merger system favored the triple point source model (SDSS J0849+1114). All of the multiple point source systems had Bayes factors that favored the same model when using both informative and non-informative priors, with the exception of SDSS J2356$-$1016, which had a Bayes factor that favored the dual point source model only when using informative priors (however, false positive tests were carried out to ensure that the Bayes factor using informative priors was strong, see Foord et al. 2020b). 
\par The posterior distributions returned by \BAYMAX{} showed that the best-fit locations of each multiple point-source system coincided with the optical nuclei of galaxies within the merger (as determined by SDSS DR16). By analyzing and fitting the individual X-ray spectra of each point source component, it was found that all point sources have unabsorbed $2-7$ keV luminosities greater than 10$^{40}$ erg s$^{-1}$, with the exception of the secondary X-ray point source in NGC 3341. Classifying X-ray point sources with $L_{\mathrm{2-7 \ keV}} > 10^{41}$ erg s$^{-1}$ as bona fide AGN (see Section~\ref{chap6:origin}), NGC 3341 was classified as a single AGN system, while SDSS J161+2352, SDSS J1708+2153, SDSS J2356$-$1016, and SDSS J0849+1114 were concluded to have at least one AGN (the primary point source in all cases). In Table~\ref{chap6:tabTripleGalInfo}, we list the X-ray results from Foord et al. (2020b).
\begin{table*}[t]
\begin{center}
\caption{Sample Properties}
\setlength\tabcolsep{2pt}
\label{chap6:tabTripleGalInfo}
\begin{tabular*}{0.95\textwidth
}{lcccccc}
	\hline
	\hline
	\multicolumn{1}{c}{Galaxy Name}  & \multicolumn{1}{c}{Redshift} &
	\multicolumn{1}{c}{$D_{L}$ (Mpc)} &  \multicolumn{1}{c}{X-ray Det.} & \multicolumn{1}{c}{$W1$ (Vega mag.)} & \multicolumn{1}{c}{$W2$ (Vega mag.)} & \multicolumn{1}{c}{$W3$ (Vega mag.)} \\
	\multicolumn{1}{c}{(1)} & \multicolumn{1}{c}{(2)} & \multicolumn{1}{c}{(3)} & \multicolumn{1}{c}{(4)} & \multicolumn{1}{c}{(5)} & \multicolumn{1}{c}{(6)} & \multicolumn{1}{c}{(7)}\\
	\hline
	SDSS J084905.51+111447.2 & 0.077 & 356.0 & Yes (\emph{p}) & 12.46$\pm$0.03 & 10.78$\pm$0.02 & 7.23$\pm$0.02   \\ 
	SDSS J084905.51+111447.2 SW & \dots & \dots &  Yes (\emph{s}) & 15.01$\pm$0.12 & 15.17$\pm$0.35 & 11.18$\pm$0.47 \\
	SDSS J084905.51+111447.2 NW & \dots & \dots &  Yes (\emph{t}) & 12.46$\pm$0.03 & 10.78$\pm$0.02 & 7.23$\pm$0.02   \\
	\hline
	SDSS J085837.67+182223.3 & 0.059 & 265.7 & No & 11.88$\pm$0.02 & 11.63$\pm$0.02 & 8.39$\pm$0.03  \\
	SDSS J085837.67+182223.3 SW & \dots & \dots &  No & 11.88$\pm$0.02 & 11.63$\pm$0.02 & 8.39$\pm$0.03  \\
	SDSS J085837.67+182223.3 SE & \dots & \dots &  No & 11.88$\pm$0.02 & 11.63$\pm$0.02 & 8.39$\pm$0.03 \\
	\hline
	SDSS J102700.40+174900.8  & 0.066 & 298.7 &  Yes (\emph{p}) & 12.06$\pm$0.04 & 11.89$\pm$0.04 & 8.46$\pm$0.05  \\	
	SDSS J102700.40+174900.8 N & \dots & \dots &  Yes (\emph{s}) & 12.06$\pm$0.04 & 11.89$\pm$0.04 & 8.46$\pm$0.05 \\
	SDSS J102700.40+174900.8 W & \dots & \dots &  No & 12.06$\pm$0.04 & 11.89$\pm$0.04 & 8.46$\pm$0.05 \\ 
    \hline
    NGC 3341  & 0.027 & 118.7 &  Yes (\emph{p}) & N/A & N/A & N/A  \\
    NGC 3341 SW & \dots & \dots &  Yes (\emph{s}) & 11.19$\pm$0.022 & 11.06$\pm$0.02 & 7.43$\pm$0.02 \\ 
    NGC 3341 NW & \dots & \dots &  No & N/A & N/A & N/A  \\
    \hline
    SDSS J163115.52+235257.5 & 0.059 & 265.7 &  Yes (\emph{p}) & $11.82\pm0.25$ & 11.03$\pm$0.02 & 7.96$\pm$0.02 \\
    SDSS J163115.52+235257.5 NE & \dots & \dots &  Yes (\emph{s}) & $11.82\pm0.25$ & 11.03$\pm$0.02 & 7.96$\pm$0.02   \\
    SDSS J163115.52+235257.5 NW & \dots & \dots &  No & N/A & N/A & N/A    \\
    \hline
    SDSS J170859.12+215308.0 & 0.072 & 327.2 &  Yes (\emph{p}) & 11.66$\pm$0.03 & 10.86$\pm$0.02 & 8.33$\pm$0.02 \\
    SDSS J170859.12+215308.0 NE & \dots & \dots &  Yes (\emph{s}) & 13.48$\pm$0.06 & 13.32$\pm$0.08 & 10.74$\pm$0.14  \\
    SDSS J170859.12+215308.0 SW & \dots & \dots &  No & 11.66$\pm$0.03 & 10.86$\pm$0.02 & 8.33$\pm$0.02  \\
    \hline
    SDSS J235654.30-101605.3 & 0.074 & 336.8 &  Yes (\emph{p}) & 10.84$\pm$0.02 & 9.81$\pm$0.02 & 6.76$\pm$0.02  \\
    SDSS J235654.30-101605.3 SE & \dots & \dots &  Yes (\emph{s}) & 10.84$\pm$0.02 & 9.81$\pm$0.02 & 6.76$\pm$0.02  \\
    SDSS J235654.30-101605.3 NE & \dots & \dots &  No & N/A & N/A & N/A  \\
    \hline
	\hline 
\end{tabular*}
\end{center}
Note. -- Columns: (1) Galaxy name; (2) spectroscopic redshift from SDSS DR16; (3) luminosity distance; (4) X-ray detection results from Foord et al. (2020b); (5) \emph{W1} magnitude measurement taken from the AllWISE Catalog, in Vega magnitudes; (6) \emph{W2} magnitude measurement taken from the AllWISE Catalog, in Vega magnitudes; (7) \emph{W3} magnitude measurement taken from the AllWISE Catalog, in Vega magnitudes. Column 4 represents whether an X-ray point source was found at the location of the galactic nucleus, and we denote whether the X-ray point source is the ``primary" (\emph{p}), ``secondary"(\emph{s}), or ``tertiary"(\emph{t}), as defined in Foord et al. (2020b). Galaxies with the same \emph{WISE} magnitudes in a given system are not individually resolved. Galaxies with ``N/A" have no \emph{WISE} coverage. 
\end{table*}

\section{Origin of X-ray Emission}
\label{chap6:origin}
In the following section we aim to identify the origin of X-ray emission for each triple merger with Bayes factors that favor multiple point source models (SDSS J0849+1114, SDSS J1027+1749, NGC 3341, SDSS J1631+2342, SDSS J1708+2143, and SDSS J2356$-$1016; see Table~\ref{chap6:tabTripleGalInfo}). Each system was found to have either 2 or 3 X-ray point sources, hereafter the ``primary", ``secondary", or ``tertiary". We incorporate IR observations of each system to better classify each X-ray point source detected by \BAYMAX{}. In particular, we compare the X-ray luminosities calculated in Foord et al. (2020b) to the X-ray emission associated with the galactic X-ray binary (XRB) population. 
\par In Foord et al. 2020b, all X-ray point sources were fit with phenomenological spectral models (i.e., described by a single or double power-law), while those with over 100 counts between $0.5-8$ keV were also fit with a physically motivated model. The latter was implemented via the {\tt BNTorus} model in XSPEC, which considers an X-ray source surrounded by a toroidal structure (see \citealt{Brightman2011} for more detail). For the systems fit with both models, we found that the phenomenological and physically-motivated models returned spectral parameters values consistent with one-another. The exception to this is the primary in NGC 3341 (NGC 3341$_{p}$, see Table~\ref{chap6:tabTripleGalInfo}) where the physically motivated model was consistent with higher levels of obscuration (as measured by X-ray spectral parameter $N_{H}$, measured in units of cm$^{-2}$) and thus a higher intrinsic X-ray luminosity. This was interpreted as a possible effect of the extremely high levels of absorption associated with the system ($> 10^{22}$ cm$^{-2}$), where a physically-motivated model describing the torus may better constrain parameters. Thus, the X-ray flux and luminosity values we quote in the following section are taken from our phenomenological spectral fits, with the exception of NGC 3341$_{p}$, where the values are taken from spectral fits using {\tt BNTorus} (and we refer the reader to Foord et al. 2020b for more details). All errors bars reported in this section are evaluated at the 99.7\% confidence level, unless otherwise stated.
\subsection{XRB Contamination}
\par We define an AGN as an X-ray point source with X-ray luminosity $L_{\mathrm{2-7 \ keV}} > 10^{40}$ erg s$^{-1}$ as well as X-ray luminosity greater than expected from X-ray contamination from galactic X-ray binaries (XRBs). Bona fide AGN are classified as point sources with unabsorbed 2$-$7 keV luminosities $L_{2-7~\mathrm{keV, unabs}}>10^{41}$ erg s$^{-1}$, while likely AGN are classified as point sources with unabsorbed 2$-$7 keV luminosities $L_{2-7~\mathrm{keV, unabs}}>10^{40}$ erg s$^{-1}$. Any point source with $L_{2-7~\mathrm{keV, unabs}}<10^{40}$ erg s$^{-1}$ is conservatively not classified as an AGN. Generally, for point sources with X-ray luminosities below $L_{2-7~\mathrm{keV, unabs}} < 10^{40}$ erg s$^{-1}$, X-ray binaries or ultraluminous X-ray sources (ULX) can explain the accretion nature. The majority of the high-mass X-ray binary (HMXB) population has $2$--$7$ keV X-ray luminosities between 10$^{38}$--10$^{39}$ erg s$^{-1}$, while the ULX population can reach even higher X-ray luminosities ($L_{\mathrm{2-7 \ keV}}>$ 10$^{39}$ erg s$^{-1}$; \citealt{Swartz2011, Walton2011}). However, their X-ray luminosity functions (XLF) have been measured to drop off sharply at $L_{\mathrm{2-7 \ keV}}=$10$^{40}$ erg s$^{-1}$ (e.g., \citealt{Mineo2012, Sazonov2017, Lehmer2019}), and previous studies have shown that the majority of nuclear X-ray point sources with $L_{\mathrm{2-7 \ keV}}>$ 10$^{40}$ erg s$^{-1}$ are highly likely to be emission associated with SMBHs \citep{Foord2017a, Lehmer2019}. 
\par However, in merging systems, amplified star formation rates can increase the surrounding X-ray emission, and the population of XRBs in mergers may indeed have different XLFs than the late- and early-type galaxies included in past studies. Thus, an important step for properly classifying each X-ray point sources with $10^{40} < L_{2-7~\mathrm{keV, unabs}} < 10^{41}$ erg s$^{-1}$, is to compare the X-ray luminosity to the expected X-ray contribution from high-mass X-ray binaries. In particular, the high-mass X-ray binary XLF traces recent star formation within the galaxy \citep{Sunyaev1978, Grimm2003, Lehmer2010, Mineo2012, Lehmer2019}; given the high X-ray luminosity of each point-source in our sample, as well as the high estimated star formation rates (SFRs $>$ 2 $M_{\odot}$ yr$^{-1}$; see Table~\ref{tab:SFR}), HMXBs should be the dominant source of contamination (with respect to low-mass X-ray binaries, whose XLF scales with the total stellar mass, $M_{\ast}$, of the galaxy). Thus, in our following analysis, we estimate the galactic X-ray luminosity expected from galactic HMXBs (but we note that the LMXB X-ray luminosity contribution for each of the systems in our sample is expected to be approximately an order of magnitude less than the HMXB contribution).
\par We estimate the total expected 2$-$7 keV luminosity from the high-mass X-ray binary population, $L^{\mathrm{gal}}_{\mathrm{HMXB}}$, by calculating the SFR of each system and using the analytical prescription presented in \cite{Lehmer2019}. Specifically, \cite{Lehmer2019} constrain the scaling relations for XRBs in a sample of 38 nearby and IR-bright galaxies by fitting a global XLF model that contains contributions from both HMXBs (which scales with the SFR) and LMXBs (which scales with $M_{\ast}$). In particular, they find $L^{\mathrm{gal}}_{\mathrm{HMXB}} = \beta\times\mathrm{SFR}$, where $\beta = 39.71^{+0.14}_{-0.09}$ ergs s$^{-1}$ ($M_{\odot}$ yr$^{-1}$)$^{-1}$.
\par Due to the dusty environments of the mergers, we estimate the SFRs of each of the mergers in our sample using IR diagnostics, which are less prone to obscuration. The IR-derived SFRs (SFR$_{\mathrm{IR}}$) are estimated using the total estimated IR luminosity ($L_{TIR}$, between $8-1000$ microns) of each system as presented in \cite{Bell2003}: $L_{TIR} = 2.52 \times 10^{-14} \times (2.54\times f_{60} + f_{100})$. Here $f_{60}$ and $f_{100}$ are the 60$\mu$m and 100$\mu$m fluxes in Jy. For each system, we use available 60 $\mu$m and 100 $\mu$m flux density values from the IRAS Faint Source Catalog \citep{Moshir1990}. The SFR$_{\mathrm{IR}}$ value is then calculated using the relation presented in \cite{Kennicutt2012}, where $\log{\mathrm{SFR_{\mathrm{IR}}}} = \log(L_{TIR}) - 43.41$. If no IRAS data are available, we use archival \emph{WISE} W3-band (12 $\mu$m) observations to estimate the 60 $\mu$m and 100 $\mu$m flux density values (outlined in further detail in Section~\ref{chap6:subsecJ1631}, see \citealt{Terrazas2016}).
\par Given the angular resolutions of IRAS (0.5$^{\prime}-2^{\prime}$) and \emph{WISE} ($\approx$6.5\arcsec at 12 $\mu$m) we are unable to estimate the \emph{individual} SFRs for each galaxy in the merger, and thus $L^{\mathrm{gal}}_{\mathrm{HMXB}}$ represents the expected X-ray luminosity from HMXBs across all three galaxies. For 1 dual X-ray point source system, where the separation between the two host galaxies are individually resolved by \emph{WISE} (SDSS J1708+2153, see below), we estimate the total $L^{\mathrm{gal}}_{\mathrm{HMXB}}$ from each galaxy individually. Even in this best-case scenario, where each galaxy has its own SFR estimation, we stress that the calculated $L^{\mathrm{gal}}_{\mathrm{HMXB}}$ is the X-ray luminosity from HMXB contribution across the entire galaxy, whereas our X-ray detections are contained within the central 2\arcsec~where the nuclear X-ray contribution from HMXB, $L^{\mathrm{nucleus}}_{\mathrm{HMXB}}$, can be an order of magnitude smaller than $L^{\mathrm{gal}}_{\mathrm{HMXB}}$ \citep{Foord2017a}.
\par Lastly, we note that \cite{Lehmer2019}  find that the variance of $\beta$ is dependent on the measured SFR, where the scatter is $\approx 0.3$ dex for SFRs $>$ 2 $M_{\odot}$ yr$^{-1}$, but grows as a function of decreasing SFR below this threshold (up to $\approx 0.7$ dex). The error bars associated with the global best-fit represent an average error across SFR space. We emphasize that none of our estimated galactic SFRs are below $2 M_{\odot}$ yr$^{-1}$, with the exception of the secondary X-ray point source in SDSS J1708+1749. In this case, we find that the errors on $\beta$, and thus $L^{\mathrm{gal}}_{\mathrm{HMXB}}$, are consisted with what is estimated for the system, given its specific SFR of the system (as outlined in table 6 in \citealt{Lehmer2019}).
\subsection{Radio Observations}
We also analyze available archival data from the VLA Faint Images of the Radio Sky at Twenty-Centimeters (FIRST) Survey \citep{White1997}. Although archival FIRST observations will resolve most of the mergers (where the resolution is $\sim$5\arcsec), it may be possible to resolve the primary and secondary galaxy in SDSS J1708+2153 (where the separation is larger than 5\arcsec), as well as estimate SFRs via a second method. The non-thermal radio luminosity provides a dust-insensitive measure of the current star formation on $\sim$100 Myr time-scales. However, radio emission is a more indirect indicator of SFRs than other wavelengths, and radio-SFR relations are mostly based on tight correlations measured between far-infrared (FIR) emission and rest-frame 1.4 GHz radio emission (e.g., \citealt{Condon1992, Yun2001}).
\par We find archival VLA FIRST observations for SDSS J0849+1114, SDSS J0858+1822, SDSS J1027+1749, SDSS J1631+2352, and SDSS J2356$-$1016. NGC 3341 has archival EVLA data, which are analyzed in \cite{Bianchi2013}. SDSS J1708+2153 has no available archival VLA data, and thus we are unable to analyze the radio emission associated with the primary and secondary galaxy. We estimate the levels of star formation using 1.4 GHz-SFR relation presented in \cite{Davies2017}. The analysis calibrated the 1.4 GHz SFRs by analyzing a combination of the Galaxy And Mass Assembly (GAMA) survey and the FIRST survey. Specifically, they compare the 1.4 GHz luminosity in comparison to measured UV and total IR SFRs to derive: $\log{\mathrm{SFR}_{1.4\mathrm{GHz}}}=0.66(\pm0.02)\times\log{L_{1.4\mathrm{GHz}}}-14.02(\pm0.39)$, where $L_{{1.4\mathrm{GHz}}}$ is the 1.4 GHz luminosity in units of erg s$^{-1}$ Hz$^{-1}$. We calculate errors on the 1.4 GHz luminosity density by adding an additional uncertainty in quadrature, taken to be 5\% of the integrated flux (see, e.g., \citealt{Perley2017}).
\par Regarding NGC 3341, \cite{Bianchi2013} list the 5 GHz luminosity densities for both the primary and secondary X-ray point source, which allows us to individually estimate the SFR$_{1.4\mathrm{GHz}}$ values of each galaxy. However, the host galaxy of the primary X-ray point source is not completely resolved from the brighter (in radio) host galaxy of the secondary X-ray point source, and therefore may contain emission from the secondary. Thus, our estimates for the SFR$_{1.4\mathrm{GHz}}$ of the primary's host-galaxy represents an upper-limit to the actual value. We convert the 5 GHz EVLA detections to 1.4 GHz, assuming $f_{\nu} \propto \nu^{-0.5}$, which has been measured for Seyfert galaxies (i.e., \citealt{Ho2001}).
\par We tabulate our estimated SFR$_{1.4\mathrm{GHz}}$ values in Table~\ref{tab:SFR}. We find that our values for SFR$_{1.4\mathrm{GHz}}$ agree with our values for SFR$_{\mathrm{IR}}$, with the exception of SDSS J2356$-$1016, where SFR$_{1.4\mathrm{GHz}}$ is significantly less than, and inconsistent within the errors of SFR$_{\mathrm{IR}}$.
\par Such a discrepancy is consistent with the findings of other studies comparing various SFRs between wavelengths, where the SFR$_{1.4\mathrm{GHz}}$ values were found to be, on average, a factor of $\sim$2.5 less than SFR$_{\mathrm{IR}}$ values (e.g., \citealt{Murphy2011}). These discrepancies are likely due to the fact that the FIR and radio emission are not necessarily co-spatial or co-temporal. The FIR emission originates from dust clouds relatively close to the young stars, and is expected to emit fairly soon after a localized burst of star formation; on the other hand, the radio emission originates from electrons traveling through the galaxy, interacting with its magnetic field, and can emit relatively far from and later than the original star formation. These differences can explain most of the scatter in the IR/radio correlations, and it is likely that the relation is dependent on the levels of recent and/or current star formation (see, e.g., \citealt{Bressan2002} where they find that the IR-to-radio luminosity ratio varies by up to an order of magnitude with the age of the starburst for a range of star formation histories). We emphasize, however, that for J2356$-$1016, SFR$_{\mathrm{IR}}$ (the value that we use to derive the XRB contamination) is the higher of the two SFRs, resulting in a higher estimated $L^{\mathrm{gal}}_{\mathrm{HMXB}}$, and thus the most conservative analysis.

\subsection{SDSS J1027+1749}
\label{chap6:subsecJ1027}
In Foord et al. (2020b), the primary and secondary X-ray point source were calculated to have a total observed $0.5$--$8$ keV flux of $5.7_{-1.6}^{+1.3} \times 10^{-15}$ erg s$^{-1}$ cm$^{-2}$ and $4.8_{-2.2}^{+1.1} \times 10^{-15}$ erg s$^{-1}$ cm$^{-2}$ s$^{-1}$, respectively. This corresponds to a rest-frame $2$--$7$ keV luminosity of $3.2_{-0.9}^{+0.8} \times 10^{40}$ erg s$^{-1}$ and $2.6_{-1.2}^{+0.6} \times 10^{40}$ erg s$^{-1}$ at $z=0.066$. 
\par The 2$-$7 keV luminosities of each point source are greater than 10$^{40}$ erg s$^{-1}$ at the 99.7\% confidence level, which can be comfortably attributed to AGN emission. In order to better understand how these luminosities compare to the expected population of high-mass X-ray binaries, we estimate the total expected 2$-$7 keV luminosity from the high-mass X-ray binary population, $L^{\mathrm{gal}}_{\mathrm{HMXB}}$. Using archival IRAS observations, we find the expected X-ray luminosity from the galactic HMXB population to be $4.7^{+1.8}_{-0.9} \times10^{40}$ erg s$^{-1}$. Although this is consistent with X-ray luminosities measured for both the primary and secondary point source source, we emphasize that $L^{\mathrm{gal}}_{\mathrm{HMXB}}$ represents the expected X-ray luminosity across all three systems, whereas our X-ray detections with \BAYMAX{} are estimates of the individual, \emph{nuclear} X-ray emission. 
\par For example, if we assume that the individual IR luminosities for each galaxy scales as the measured SDSS $i$-band flux ratios,  we can estimate the fractional value of $L^{\mathrm{gal}}_{\mathrm{HMXB}}$ associated with each galaxy. We analyze the SDSS $i$-band corrected frame for SDSS J1027+1749 and measure the flux within circular regions (with 1$\arcsec$ radii) centered on the locations of SDSS J1027+1749, SDSS J1027+1749 N, and SDSS J1027+1749 W (given that the IRAS observations, and thus calculated SFRs, include all three galaxy nuclei). We compare the flux within each region to the combined flux from the three galaxies. Our errors on the flux within each region are estimated using Gaussian statistics. We find the SDSS J1027+1749 and SDSS J1027+1749 N contain 25.6 $\pm$ 5.9\% and 35.6 $\pm$ 8.1\% of the $i$-abnd emission, corresponding to fractional $L^{\mathrm{gal}}_{\mathrm{HMXB}}$ values of $1.2_{-0.5}^{+0.4}\times10^{40}$ erg s$^{-1}$ and $1.7_{-0.8}^{+0.5}\times10^{40}$ erg s$^{-1}$, respectively. Not only are these values inconsistent with the X-ray luminosities of the primary and secondary point sources in  SDSS J1027+1749 at the 95\% confidence level, but these fractional $L^{\mathrm{gal}}_{\mathrm{HMXB}}$ values represent an estimation for the total galactic X-ray luminosity. As such, we conclude that the triple merger is likely composed of 2 X-ray AGN.

\subsection{NGC 3341}
\label{chap6:subsecN3341}
For the primary, Foord et al. (2020b) calculated a total observed $0.5$--$8$ keV flux of $1.39_{-0.01}^{+0.01} \times 10^{-12}$ erg s$^{-1}$ cm$^{-2}$, while the secondary was calculated to have a total observed $0.5$--$8$ keV flux of $2.7_{-0.8}^{+0.6} \times 10^{-15}$ erg s$^{-1}$ cm$^{-2}$ s$^{-1}$. This corresponds to a rest-frame $2$--$7$ keV luminosity of $1.18_{-0.01}^{+0.02} \times 10^{42}$ erg s$^{-1}$ and $2.3^{+0.9}_{-0.7} \times 10^{39}$ erg s$^{-1}$ at $z=0.027$. 
\par NGC 3341 is not included in the IRAS All-Sky Survey, and the existing archival \emph{WISE} observation is centered on the SW nucleus. However, given the primary point source's relatively high X-ray luminosity, and that standard diagnostics based on lines ratio unequivocally indicate that it is a Seyfert 2 galaxy \citep{Barth2008, Bianchi2013}, we conclude that the primary galaxy hosts an AGN. Futhermore, using our estimated value of SFR$_{1.4\mathrm{GHz}}$, in place of SFR$_{\mathrm{IR}}$, we estimate $L^{\mathrm{gal}}_{\mathrm{HMXB}}<0.2 \times 10^{39}$ erg s$^{-1}$, two order of magnitudes smaller than what we measure for the $2$--$7$ keV luminosity of the primary point source. Given that the secondary does not meet our AGN X-ray luminosity criterion ($L_{\mathrm{2-7 \ keV}}<10^{40}$ erg s$^{-1}$), we conclude that the triple merger system is composed of a single AGN. 

\subsection{SDSS J1631+2352}
\label{chap6:subsecJ1631}
Foord et al. (2020b) calculated a total observed $0.5$--$8$ keV flux of $2.29^{+0.01}_{-0.01} \times 10^{-12}$ erg s$^{-1}$ cm$^{-2}$, and $5.3^{+4.8}_{-3.9} \times 10^{-15}$ erg s$^{-1}$ cm$^{-2}$ for the primary and secondary, respectively. They correspond to rest-frame $2$--$7$ keV luminosities, at $z=0.059$, of $1.33^{+0.01}_{-0.01} \times 10^{43}$ erg s$^{-1}$ and $2.5^{+2.3}_{-1.8} \times 10^{40}$ erg s$^{-1}$. 
\par Both X-ray luminosities are above 10$^{40}$ erg s$^{-1}$, while given the primary's 2$-$7 keV luminosity, we categorize it as a bona fide AGN. We calculate $L^{\mathrm{gal}}_{\mathrm{HMXB}}$ using IR \emph{WISE} W3 band (12 $\mu$m) observations. To convert the 12 $\mu$m flux to the equivalent FIR IRAS bands, we use the $f_{60\mu\mathrm{m}}/f_{12\mu\mathrm{m}}$, $f_{60\mu\mathrm{m}}/f_{25\mu\mathrm{m}}$, and $f_{60\mu\mathrm{m}}/f_{100\mu\mathrm{m}}$ flux ratios presented in \cite{Terrazas2016}. To ensure the validity of this approach, we compare the estimated FIR IRAS-band flux values using the \emph{WISE} W3 band detections to the actual IRAS-band flux values, for the triple mergers in our sample with both IRAS and $WISE$ observations. We find that the estimated FIR IRAS-band flux values are consistent with the actual IRAS-band flux values. 

\par For the primary and NE galaxy, we find $L^{\mathrm{gal}}_{\mathrm{HMXB}}=1.5_{-0.3}^{+0.6}\times10^{40}$ erg s$^{-1}$, which is consistent with the measured X-ray luminosity of secondary, at the 99.7\% C.L. Similar to SDSS J1027+1749, if we assume that the individual IR luminosities for SDSS J1631+2352 and SDSS J1631+2352 NE scale as the measured $i$-band flux ratios (62.2 $\pm$ 18.9\% and 37.8 $\pm$ 11.9 \%, respectively), we estimate fractional values of $L^{\mathrm{NE~gal}}_{\mathrm{HMXB}}<10^{40}$ erg $s^{-1}$ at the 99.7\% confidence level.
\par As part of the BAT X-ray survey, the total IR luminosity of the merger was also estimated \cite{Ichikawa2019}. The analysis presented in \cite{Ichikawa2019} quantifies the luminosity contribution of the AGN to the total IR emission (defined as the emission between 5$-$1000 $\mu$m, and thus is a good comparison to our values of $L_{\mathrm{IR}}$) by decomposing the IR spectral energy densities into an AGN and starburst component (via templates for an AGN torus and a star-forming galaxy). The results from their analysis allows us to isolate the expected IR luminosity from star formation, for the full merger. We note that the total IR luminosity they estimate for the system ($L_{IR} = 2.4 \times 10^{44}$ erg s$^{-1}$) is consistent with what we estimate ($L_{IR} = 2.3 \pm 1.8 \times 10^{44}$ erg s$^{-1}$) using a less thorough method. Using their estimate for the expected IR luminosity from star formation, for the full merger, we find that the X-ray contribution from HMXBs is estimated to be $8.6^{+3.3}_{-1.6} \times 10^{39}$ erg s$^{-1}$, inconsistent with the X-ray luminosity measured for the secondary X-ray point source at the 99.7\% confidence level.
\par Given that our two different approaches -- using the $i$-band flux ratios as well as the isolated $L_{TIR}$ emission from star formation -- predict $L^{\mathrm{NE~gal}}_{\mathrm{HMXB}}$ that are inconsistent with what we measure for SDSS J1631$_{s}$, we conclude that the emission detected from the secondary point source can be confidently attributed to an AGN. For both SDSS J1631+2352 and SDSS J1027+1729, future observations with IFU spectroscopy will allow for a more detailed analysis of the individual SFRs of each galaxy in the merger systems.
\subsection{SDSS J1708+2153}
\label{chap6:subsecJ1708}
For the primary, Foord et al. (2020b) calculated a total observed $0.5$--$8$ keV flux of $1.46^{+0.01}_{-0.01} \times 10^{-12}$ erg s$^{-1}$ cm$^{-2}$, while the secondary has a total observed $0.5$--$8$ keV flux of $4.6^{+1.6}_{-0.2} \times 10^{-15}$ erg s$^{-1}$ cm$^{-2}$ s$^{-1}$. The measured flux values correspond to rest-frame $2$--$7$ keV luminosities of $1.17^{+0.01}_{-0.01} \times 10^{43}$ erg s$^{-1}$ and $3.5^{+0.6}_{-1.2} \times 10^{40}$ erg s$^{-1}$ at $z=0.072$. %
\par Both the primary and second point sources have $2$--$7$ keV luminosities above 10$^{40}$ erg s$^{-1}$ at the 99.7\% confidence level. Due to the larger angular separation between the two systems ($\approx$6.59\arcsec), we use resolved IR \emph{WISE} W3 band (12 $\mu$m) observations of each galaxy to estimate their individual $L^{\mathrm{gal}}_{\mathrm{HMXB}}$. For the primary and NE galaxy, we find $L^{\mathrm{pri~gal}}_{\mathrm{HMXB}}=1.6_{-0.3}^{+0.6}\times10^{40}$ and $L^{\mathrm{NE~gal}}_{\mathrm{HMXB}}=0.2_{-0.1}^{+0.1}\times10^{40}$, inconsistent with the measured X-ray luminosities of the primary (at the 99.7\% C.L.) and secondary (at the 95\% C.L.) point source. 
\par As part of the BAT X-ray survey, the total IR luminosity of the merger was also estimated \cite{Ichikawa2019}. We note that the total IR luminosity they estimate for the system ($L_{IR} = 1.2 \times 10^{44}$ erg s$^{-1}$) is lower than what we estimate ($L_{IR} = 2.4 \pm 0.8 \times 10^{44}$ erg s$^{-1}$). Using their estimate for the expected IR luminosity from star formation, for the full merger, we find that the X-ray contribution from HMXBs is estimated to be less than 10$^{40}$ erg s$^{-1}$ at the 99.7\% confidence level. Using our various approaches to estimate the X-ray luminosity from HMXBs in each galaxy, we can safely classify the two X-ray point sources in SDSS J1708+2153 as AGN.

\subsection{SDSS J2356$-$1016}
\label{chap6:subsecJ2356}
The primary was found to have a total observed $0.5$--$8$ keV flux of $1.60^{+0.01}_{-0.01} \times 10^{-12}$ erg s$^{-1}$ cm$^{-2}$, while the secondary has a total observed $0.5$--$8$ keV flux of $7.3^{+7.0}_{-3.5} \times 10^{-15}$ erg s$^{-1}$ cm$^{-2}$ s$^{-1}$. This corresponds to a rest-frame $2$--$7$ keV luminosity of $3.11^{+0.04}_{-0.04} \times 10^{43}$ erg s$^{-1}$ and $8.0^{+9.6}_{-1.1} \times 10^{40}$ erg s$^{-1}$ at $z=0.074$. The spectral fit of the primary point source shows relatively high (with respect to the other dual X-ray point sources in the sample) levels of absorption, with $N_{H}=7.83_{-0.14}^{+0.18}\times10^{22}$ cm$^{-2}$.
\par Both X-ray point sources have $2$--$7$ keV luminosities greater than 10$^{40}$ erg s$^{-1}$ at the 99.7\% confidence level. SDSS J2356$-$1016 has no IRAS observations, and thus we use IR \emph{WISE} W3 band observations of the entire triple merger system to estimate $L^{\mathrm{gal}}_{\mathrm{HMXB}}$. Following the procedure outline in Section~\ref{chap6:subsecJ1708}, we estimate the total HMXB X-ray contribution of the triple merger system to be $L^{\mathrm{gal}}_{\mathrm{HMXB}} = 7.1^{+2.7}_{-1.3}\times10^{40}$ erg s$^{-1}$. Although the primary X-ray point source is well above this luminosity, the secondary X-ray point source is consistent at the 99.7\% C.L. with $L^{\mathrm{gal}}_{\mathrm{HMXB}}$. Thus, we aim to better understand the fractional contribution of $L^{\mathrm{gal}}_{\mathrm{HMXB}}$ to just the SE galaxy. \cite{Pfeifle2019a} calculated SFRs of each nucleus individually using the estimated, intrinsic, H$\alpha$ line fluxes. These values were estimated from individually resolved Pa$\alpha$ emission-lines detected via near-IR longslit spectra of each nucleus from the Large Binocular Telescope (LBT). They find a SFR of 3.82 $M_{\odot}$ yr$^{-1}$ for the SE nucleus, corresponding to $L^{\mathrm{gal}}_{\mathrm{HMXB}} = 9.9_{-1.9}^{+3.8}\times10^{39}$ erg s$^{-1}$. This value for the expected galactic HMXB X-ray emission for SDSS J2356$-$1016 SE is inconsistent with what we measure for J2356$-$1016$_{s}$, at the 99.7\% Thus, we classify both point sources as AGN.
\subsection{SDSS J0849+1114}
The total $0.5$--$8$ keV fluxes of the primary, secondary, and tertiary X-ray point sources were found to be: $6.54^{+0.28}_{-0.22} \times 10^{-14}$ erg s$^{-1}$ cm$^{-2}$, $4.0^{+1.6}_{-1.3} \times 10^{-15}$ erg s$^{-1}$ cm$^{-2}$, and $3.2^{+1.0}_{-1.3} \times 10^{-15}$ erg s$^{-1}$ cm$^{-2}$. This corresponds to rest-frame $2$--$7$ keV luminosities (at $z=0.059$) of: $1.60^{+.80}_{-.40} \times 10^{42}$ erg s$^{-1}$ for the primary, $1.9^{+0.6}_{-0.7} \times 10^{40}$ erg s$^{-1}$ for the secondary, and $1.3^{+0.6}_{-0.5} \times 10^{40}$ erg s$^{-1}$ for the tertiary.
\par The nature of the three X-ray point sources has been extensively discussed in both \cite{Liu2019} and \cite{Pfeifle2019b}. For the purposes of uniformity in our analysis, we use archival IRAS observations of the triple merger to estimate the total X-ray contribution from the galactic HMXB population and find $L^{\mathrm{gal}}_{\mathrm{HMXB}}=3.5^{+1.4}_{-0.7}\times10^{40}$ erg s$^{-1}$. However, analyses carried out in \cite{Liu2019} and \cite{Pfeifle2019b} are more detailed than our approach. These past analyses allow us to check the integrity of our far-IR-based SFR values by comparing our results to those that were estimated using different approaches. For example, \cite{Liu2019} estimate the SFR SDSS J0849+1114 using \emph{HST}-measured $u$-band luminosities of each nucleus. They estimate the intrinsic $u$-band luminosity density ($L_{u}$) by correcting for internal dust extinction, and adapt the empirical SFR-$L_{u}$ calibration from \cite{Hopkins2003}. On the other hand, \cite{Pfeifle2019b} calculate the SFRs of each nucleus individually using the estimated, intrinsic, H$\alpha$ line fluxes (estimated via individually resolved Pa$\alpha$ emission-lines). We find that the total SFR that we estimate for SDSS J084+1114 (13.6 $\pm$ 1.0 $M_{\odot}$ yr$^{-1}$, see Table~\ref{tab:SFR}), is similar to, and consistent with, the results from \cite{Liu2019} ($\approx$ 17 $M_{\odot}$ yr$^{-1}$) and \cite{Pfeifle2019b} ($\approx$ 14.95 $M_{\odot}$ yr$^{-1}$)). 
\par Because the analyses carried out in \cite{Liu2019} and \cite{Pfeifle2019b} are more detailed than our approach, we use their results to better understand the origin of emission for each X-ray point source detected in SDSS J0849+1114. \cite{Liu2019} analyze radio, optical, and X-ray observations of the triple merger to best diagnose the accretion nature of each X-ray point source. Nuclear (within a 1\arcsec~radius centered on each galaxy nucleus) star formation rates were estimated using dust-corrected $U$-band $HST$ observations, and the X-ray emission of all three sources are greater than the expected X-ray emission of the nuclear HMXB population. This agrees with the findings in \cite{Pfeifle2019b}, where using measured Pa$\alpha$ emission from each nuclei, they estimate SFRs for each galaxy that result in an X-ray HMXB contribution an order of magnitude lower than the X-ray luminosity measured for each X-ray point source. Additional analyses in \cite{Liu2019} strengthen the evidence that each nucleus hosts an AGN: the primary and tertiary are detected as compact radio sources by the VLA in 9.0 GHz, and diagnostic emission-line ratios for all three nuclei (via long-slit spectroscopic observations using DIS on the Apache Point Observatory 3.5 m telescope) classify each as a type 2 Seyfert. Thus, we classify each X-ray point source as an AGN, and conclude that SDSS J0849+1114 is a triple merger with 3 AGN.
\subsection{SDSS J0858+1822: Evidence for Compton Thick AGN}
\label{chap6:subsecJ0858}
Foord et al. 2020b found that the emission of SDSS J0858+1822 was consistent with a single point source that was offset from the primary galaxy. These results agree with the results of the optical analysis presented in \cite{Husemann2020}, where the [\ion{O}{3}] emission-line peak was found to be offset by 1\arcsec~from the position of the primary galactic nucleus. The spectral realizations of the point source were best-fit with $m_{\mathrm{phen,1}}$ where $\Gamma$ was free to vary. The total observed $0.5$--$8$ keV flux was estimated to be $3.0^{+2.6}_{-1.3} \times 10^{-15}$ erg s$^{-1}$ cm$^{-2}$, corresponding to a rest-frame $2$--$7$ keV luminosity of $2.1^{+0.1}_{-0.1} \times 10^{39}$ erg s$^{-1}$ at $z=0.077$. The region is extremely soft, with best-fit $\Gamma=8.5_{-5.4}^{+0.7}$ and $\emph{HR}=-0.9_{-0.1}^{+0.4}$. These results are well-described by a concentrated region (or, clump) of the surrounding diffuse emission. However, the very soft emission can also possibly originate from AGN if the system is Compton thick, where the unobscured X-ray luminosity can be $\sim$60 times higher with respect to what is observed \citep{Lamastra2009, Marinucci2012}.
\par Using available \emph{IRAS} and VLA FIRST data, we estimate the levels of expected star formation from the system, and compare the expected $L^{\mathrm{gal}}_{\mathrm{HMXB}}$ value with what we measure in the \emph{Chandra} observation. This comparison allows for better insight on whether obscuration may be impacting the X-ray emission as observed by \emph{Chandra}. We estimate a SFR$_{\mathrm{IR}}$ for the entire merger of $\approx$10.8 $\pm$ 1.1 $M_{\odot}$ yr$^{-1}$, corresponding to an expected $L^{\mathrm{gal}}_{\mathrm{HMXB}} \approx 2.8_{-0.5}^{+1.0} \times 10^{40}$ erg s$^{-1}$. This value is almost a factor of 10 larger than the 2$-$8 keV luminosity we measure in the \emph{Chandra} observations. This may be evidence for a high-levels of gas within the merger.
\par Most recently, the optical spectrum of the secondary nucleus in SDSS J0858+1822 was found to reflect a non-AGN nature, dominated by star formation, while the primary nucleus has an optical spectrum that shows evidence for an AGN (reclassifying the system from a dual AGN, as determined using SDSS fiber spectroscopy, to a single AGN, as determined via long-slit observations, see Section~\ref{chap6:discussion}; \citealt{Husemann2020}). Follow-up observations in the hard X-rays with \emph{NuSTAR} (which are less susceptible to obscuration in Compton-thick environments), or with \emph{XMM} (which has higher sensitivity to hard X-rays than \emph{Chandra}) well as high-resolution radio follow-up, can better diagnose the true AGN activity in SDSS J0858+1822. \\
\par Combining our X-ray, IR, and radio analyses, we conclude that 1 triple merger system in our sample is composed of a single AGN: NGC 3341; 4 triple merger systems in our sample are strong dual AGN candidates, all of which are new discoveries: SDSS J1027+1749, SDSS J1631+2352, SDSS J1708+2153, and SDSS J2356$-$1016; we confirm one triple AGN system, SDSS J0849+1114; and 1 triple merger in our sample is not yet classifiable: SDSS J0858+1822.

\begin{table*}
\begin{center}
\caption{HMXB Contamination}
\label{tab:SFR}
\setlength\tabcolsep{2pt}
\begin{tabular*}{\textwidth
}{lcccccc}
	\hline
	\hline
	\multicolumn{1}{c}{Galaxy Name} &  \multicolumn{1}{c}{SFR$_{\mathrm{IR}}$ ($M_{\odot}$ yr$^{-1}$)} & \multicolumn{1}{c}{SFR$_{1.4\mathrm{GHz}}$ ($M_{\odot}$ yr$^{-1}$)} &
	\multicolumn{1}{c}{$L^{\mathrm{gal}}_{\mathrm{HMXB}}$ (10$^{40}$ erg s$^{-1}$)} &
	\multicolumn{1}{c}{sSFR (yr$^{-1}$)} &  \multicolumn{1}{c}{$L_{\mathrm{nucleus}}$ (10$^{40}$ erg s$^{-1}$)} \\
	\multicolumn{1}{c}{(1)} & \multicolumn{1}{c}{(2)} & \multicolumn{1}{c}{(3)} & \multicolumn{1}{c}{(4)} &
	\multicolumn{1}{c}{(5)} & \multicolumn{1}{c}{(6)} \\
	\hline
    SDSS J0849+1114  & 13.6 $\pm$ 1.0 & 38 $\pm$ 14 & 3.5$^{+1.4}_{-0.7}$ & $-8.8 \pm 0.1$ & $160^{+80}_{-40}$ \\ 
	SDSS J0849+1114 NW & \dots & \dots & \dots & \dots & $1.9^{+0.6}_{-0.7}$  \\ 
	SDSS J0849+1114 SW & \dots & \dots & \dots & \dots &  $1.3^{+0.6}_{-0.5}$ \\ 
	\hline 
	SDSS J0858+1822 & $10.8 \pm 1.1$ & $8.8 \pm 1.7$ & 2.8$^{+1.0}_{-0.5}$ & $-9.6 \pm 0.1$
 & $0.21_{-0.01}^{+0.01}$ \\ 
    \hline
	SDSS J1027+1749 & $18.2 \pm 1.3$ & $17 \pm 4$ & 4.7$^{+1.8}_{-0.9}$ & $-9.4 \pm 0.1$
 & $3.2_{-0.9}^{+0.8}$ \\ 
	SDSS J1027+1749 N & \dots & \dots & \dots & \dots & $2.6_{-1.2}^{+0.6}$ \\ 
	\hline
	NGC 3341 & N/A & $<0.6 \pm 0.2$ & $<0.2_{-0.1}^{+0.1}$ & $<-9.3 \pm 0.1$ & $118.2_{-1.3}^{+1.6}$ \\
	NGC 3341 SW & N/A & $2.6 \pm 0.2$ & $0.7_{-0.1}^{+0.3}$ & $-9.7 \pm 0.1$ & $0.23_{-0.07}^{+0.09}$ \\
	\hline 
	SDSS J1631+2352 & $5.8 \pm 0.1$ & $6.9 \pm 1.2$ & $1.5_{-0.3}^{+0.6}$ & $-9.6 \pm 0.1$ & $1.33^{+0.01}_{-0.01}\times10^{3}$  \\
	SDSS J1631+2352 NE & \dots & \dots & \dots & \dots & $2.5^{+2.3}_{-1.8}$ \\ 
	\hline 
	SDSS J1708+2153 & $6.1 \pm 0.1$ & -- & $1.6_{-0.3}^{+0.6}$ & $-9.7 \pm 0.1$ & $1.17_{-0.01}^{+0.01}\times10^{3}$  \\ 
	SDSS J1708+2153  NE & $0.7 \pm 0.1$ & -- & $0.2^{+0.1}_{-0.1}$ & $-10.5 \pm 0.15$ & $3.5_{-1.2}^{+0.6}$  \\ 
	\hline
	SDSS J2356$-$1016 & $27.3 \pm 1.5$ & $5 \pm 1$
    & $7.1^{+2.7}_{-1.3}$ & $-9.4 \pm 0.1$ & $3.11^{+0.04}_{-0.04}\times10^{3}$  \\ 
	SDSS J2356$-$1016 SE & \dots &\dots &\dots & \dots & $8^{+10}_{-1}$ \\  
	\hline 
	\hline
\end{tabular*}
\end{center}
Note. -- Columns: (1) Galaxy name; (2) SFR$_{\mathrm{IR}}$, in units of $M_{\odot}$ yr$^{-1}$; (3) SFR$_{1.4\mathrm{GHz}}$, in units of $M_{\odot}$ yr$^{-1}$; (4) estimated X-ray luminosity expected from HMXBs across the entire merger using SFR$_{\mathrm{IR}}$, in units of $10^{40}$ erg s$^{-1}$; (5) specific star formation rate (i.e., SFR per unit solar mass), in units of yr$^{-1}$; (6) X-ray luminosity for each X-ray point source, as analyzed in Foord et al. 2020b, in units of $10^{40}$ erg s$^{-1}$. We note that for SDSS J1708+2153 we are able to estimate SFRs and $L^{\mathrm{gal}}_{\mathrm{HMXB}}$ for each galaxy individually, given their separation and the angular resolution of \emph{WISE}. Furthermore, for our analysis on the HMXB contamination in SDSS J1027+1749 and SDSS J1631+2352, we estimate the fractional values of $L^{\mathrm{gal}}_{\mathrm{HMXB}}$ for each galaxy, using the SDSS $i$-band flux ratios in Section~\ref{chap6:origin}. All sSFRs were calculated using mass estimates following the relations presented in \cite{Zibetti2009}: $\log{M_{\ast}/M_{\odot}} = \log{L_{K}/L_{K,\odot} - 1.321 + 0.754(g-i)}$. Here, we use archival 2MASS data to estimate the total $L_{K}$ of the merger, and SDSS observations for the $g$ and $i$ magnitudes.\\
$\dagger$: Here we use the X-ray luminosity as estimated using a physically-motivated model with {\tt BNTorus}. See Section~\ref{chap6:origin} and Foord et al. 2020b for more details. \\
\end{table*}

\section{Discussion}
\label{chap6:discussion}
In the following section, we combine our results from the full sample of triple mergers to better understand the environments surrounding each AGN. We first combine our \emph{Chandra} analysis with mid-IR \emph{WISE} observations to gain insight on the preferential environments of multiple AGN systems. We then compare our X-ray classifications to optical AGN diagnostics using available optical spectroscopic observations.

\subsection{X-ray and IR AGN diagnostics}
\begin{figure}
\centering
    \includegraphics[width=0.48\textwidth]{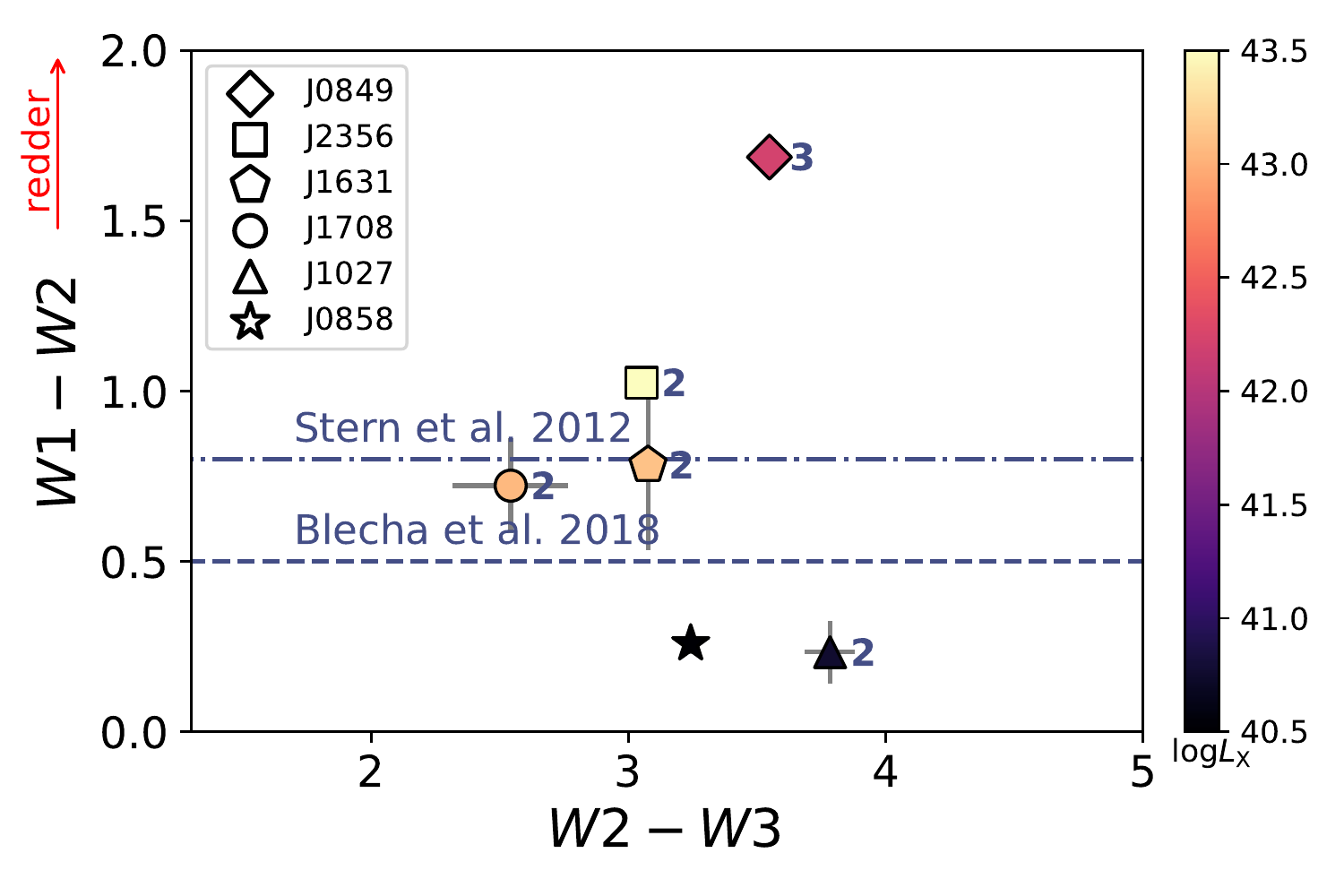}
    \includegraphics[width=0.48\textwidth]{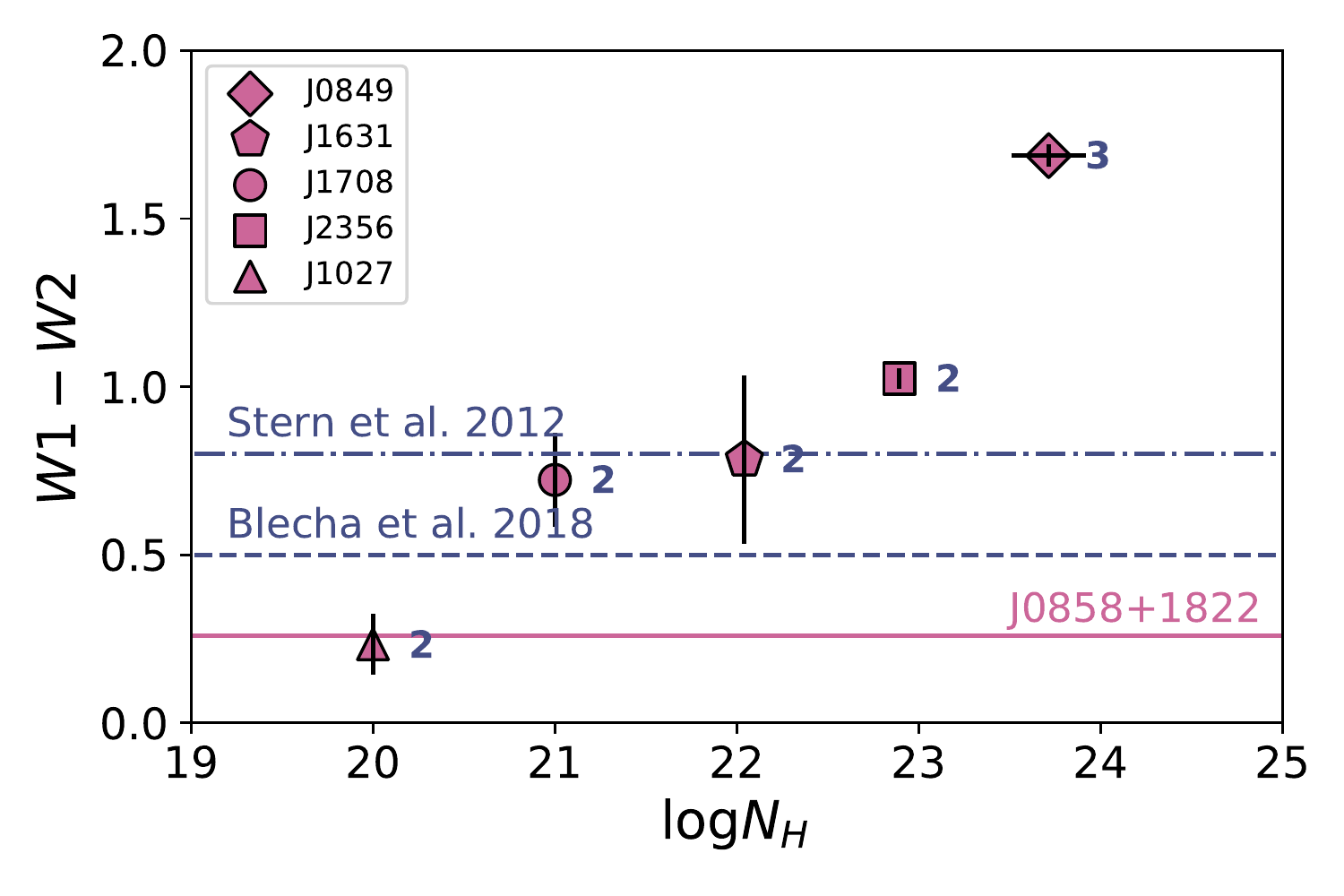}
    \caption{$W1-W2$ versus $W2-W3$ colors (\emph{top}) and $W1-W2$ versus $N_{H}$ associated with the primary AGN (\emph{bottom}) for SDSS J0849+1114 (diamond), SDSS J1027+1749 (triangle), SDSS J1631+2352 (pentagon), SDSS J1708+2153 (circle), SDSS J2356$-$1016 (square) and SDSS J0858+1822 (star on top panel). We also denote how many X-ray identified AGN are associated with each merger via numbers next to each marker. In the top panel, we find that 4 of 6 systems are identified as AGN via mid-IR color-color diagnostics, and that their $W1-W2$ colors are not correlated with the total AGN X-ray luminosity. In the bottom panel, we find a trend of increasing $N_{H}$ (associated with the primary) as a function of increasing $W1-W2$. Various AGN color-cuts, defined for single AGN by \cite{Stern2005}, and multiple AGN by \cite{Blecha2018}, as shown in dash-dot and dashed blue lines. We find that the one triple AGN system in our sample (diamond) has the highest levels of $N_{H}$ and $W1-W2$ colors, while the dual AGN (purple circles) all have lower levels. We plot the possible placements of SDSS J0858+1822 assuming a wide range of $N_{H}$ values (pink line); using our measured trend as a diagnostic for the number of AGN (with the caveat that the scatter of this relation is unconstrained until the sample size greatly increases), we may speculate that SDSS J0858+1822 has fewer than 3 AGN.}

\label{chap6:WISE}
\end{figure}

Our sample of X-ray observed triple merger systems gives us the unique opportunity to probe if and how the environments between single, dual, and triple AGN differ. Given that each of these triple merger systems are similarly separated on kilo-parsec scales, we are interested in (i) whether other environmental parameters differ between each system, and (ii) whether these differences are linked to the SMBH activity. 
\par We analyze how the \emph{WISE} $W1-W2$ colors vary as a function of the extragalactic hydrogen column density of the primary AGN, $N_{H}$. The values for $N_{H}$ are taken from the X-ray spectral fits carried out in Foord et al. (2020b). \emph{WISE} color-color diagrams ($W2-W3$ versus $W1-W2$) are a standard diagnostic used to find obscured, luminous, AGN (e.g., \citealt{Stern2005}). AGN spectra are expected to be redder than star-forming galaxies between 1-10 $\mu$m, and as a result, sit in different locations on mid-IR color-color diagrams than inactive nuclei. Various color-cuts have been defined in the literature to define emission consistent with an AGN \citep{Jarrett2011, Stern2005}, and simulations of galaxy mergers are beginning to better identify the positions of \emph{multiple} AGN systems on mid-IR color-color diagrams \citep{Blecha2018}. 
\par In the top panel of Figure~\ref{chap6:WISE} we show the mid-IR color-color plot for our sample of triple merger galaxies, excluding NGC 3341 (as the \emph{WISE} aperture only includes the NW galaxy, where we find no emission consistent with an AGN). We note that the \emph{WISE} colors account for all three galactic nuclei in all systems except SDSS J2356$-$1016 and SDSS J1631+2352, where the \emph{WISE} aperture only covers the primary and secondary galaxy (however, the tertiary galaxies excluded in both cases were found to have no X-ray emission consistent with AGN). Thus, the numbers next to each marker represent the number of X-ray detected AGN within the \emph{WISE} footprint. We find that 4 of 6 systems are identified as AGN via mid-IR color-color diagnostics, and that their $W1-W2$ colors are not correlated with the total AGN X-ray luminosity. Generally, these mid-IR color diagnostics are most useful for high-Eddington and luminous ($L_{\mathrm{X}} > 10^{43}$) AGN; lower luminosity AGN ($L_{\mathrm{2-7 \ keV}} < 10^{43}$ erg s$^{-1}$) are likely to have WISE colors contaminated by star formation (see, e.g., \citealt{Ichikawa2017}). Accurate mass measurements of each system may give more insight on possible trends between the X-ray measurements and $W1-W2$ colors.
\par We investigate a possible relation between the levels of gas obscuration and dust obscuration. In the bottom panel of Figure~\ref{chap6:WISE} we plot the measured $N_{H}$ values from the primary AGN versus the measured $W1-W2$ colors of the system. We find a general trend where the $N_{H}$ increases as a function of measured $W1-W2$. This is likely reflecting how the motions of gas and dust are coupled in merging environments, where large amounts can be siphoned into the active central region, and enshroud the central AGN. In particular, tidal forces between galaxies during mergers can cause gas and dust to be subject to substantial gravitational torques, where substantial amounts can be funneled towards the central SMBH \citep{Barnes&Hernquist1991, DiMatteo2008, Angles-Alcazar2017}. This trend agrees with recent results from \cite{Blecha2018}, where they studied the evolution of AGN during galaxy mergers via hydrodynamic simulations. They found that the levels of estimated line-of-sight gas column densities and \emph{WISE} $W1-W2$ colors tend to vary together; in particular, both parameters peak during periods of the merger such as the first pericenter passage and final coalescence.
\par Interestingly, we find that the one triple AGN system in our sample has the highest levels of $N_{H}$ and $W1-W2$ colors, while the dual AGN all have lower levels. Given that SMBHs grow, and ignite to AGN, through the accretion of cool gas, it is possible that triple merger systems with higher levels of nuclear gas (measured by $N_{H}$) will have more AGN. Investigating whether or not this trend varies with total number of AGN in a given merger will require a larger sample of multiple AGN systems.
\par Given the trend shown in the bottom panel of Figure~\ref{chap6:WISE}, and assuming that SDSS J0858+1822 is a Compton-thick system composed of one or more AGN, we may expect the placement of the merger on the $N_{H}$ versus $W1-W2$ plot will depend on the number of AGN. In particular, if the merger does indeed have three Compton-thick AGN, the $W1-W2$ color may be larger than what one would expect for 2 or 1 AGN. Because it has been shown that Compton-thick sources may appear to have low-levels of absorption if using simple models and/or low-count data sets \citep{Risaliti1999}, we plot the $N_{H}$ versus $W1-W2$ relation for SDSS J0858+1822, assuming a large range of $N_{H}$ values ($19 < \log{N_{H}} < 25$). Using our measured trend as a diagnostic for the number of AGN (with the caveat that the scatter of this relation is unconstrained until the sample size greatly increases), we may speculate that SDSS J0858+1822 has fewer than 3 AGN.

\subsection{X-ray and optical AGN diagnostics}
\label{subsection:BPT}
Due to the uniform SDSS DR16 coverage of all triple galaxies mergers in our sample, we look at the BPT diagram classifications (``Baldwin, Phillips \& Telervich", \citealt{Baldwin1981}) as defined by the [\ion{O}{3}]/H$\beta$, [\ion{S}{2}]/H$\alpha$, and [\ion{N}{2}]/H$\alpha$ narrow emission-line ratios. BPT diagrams are a diagnostic used to classify the likely ionization mechanism of hot gas in galactic nuclei; although this technique is most effective at correctly classifying for high-Eddington systems. Thus, any X-ray classified AGN that is misclassifed by BPT diagnostics may reflect certain properties of the accretion nature and/or environment of the AGN.

\begin{table*}
\begin{center}
\caption{Optical Narrow Line Classifications}
\label{tab:BPT}
\setlength\tabcolsep{2pt}
\begin{tabular*}{\textwidth
}{lcccccc}
	\hline
	\hline
	\multicolumn{1}{c}{Galaxy Name} &  \multicolumn{1}{c}{$\log{}$[\ion{N}{2}]/H$\alpha$} &
	\multicolumn{1}{c}{$\log{}$[\ion{S}{2}]/H$\alpha$} &
	\multicolumn{1}{c}{$\log{}$[\ion{O}{3}]/H$\beta$} &  \multicolumn{1}{c}{Optical Class.} & \multicolumn{1}{c}{X-ray Class.} &
	\multicolumn{1}{c}{Reference} \\
	\multicolumn{1}{c}{(1)} & \multicolumn{1}{c}{(2)} & \multicolumn{1}{c}{(3)} & \multicolumn{1}{c}{(4)} & \multicolumn{1}{c}{(5)} & \multicolumn{1}{c}{(6)} & \multicolumn{1}{c}{(7)} \\
	\hline
    SDSS J0849+1114 & $-$0.209$\pm$0.003 & $-$0.39$\pm$0.01 & 1.029$\pm$0.009 & AGN & AGN & \cite{Pfeifle2019b}\\ 
	SDSS J0849+1114 NW & $-$0.232$\pm$0.03 & $-$0.31$\pm$0.07 & 0.55$\pm$0.09 & AGN & AGN & \cite{Pfeifle2019b} \\ 
	SDSS J0849+1114 SW & $-$0.23$\pm$0.06 & $-$0.5$\pm$0.5 & 0.4$\pm$0.3 & LINER & AGN &  \cite{Pfeifle2019b} \\ 
	\hline 
	SDSS J0858+1822 & 0.070$\pm$0.006 & -- & 0.796$\pm$0.054 & AGN & N.D. & \cite{Husemann2020} \\ 
	SDSS J0858+1822 SW & $-0.301\pm$0.053 & -- & $-0.097\pm0.204$ & Composite & N.D. & \cite{Husemann2020} \\ 
	\hline 
	SDSS J1027+1749 & 0.086$\pm$0.004 & $-$0.092$\pm$0.01 & 0.505$\pm$0.081 & AGN & AGN & \cite{Liu2011b}\\ 
	SDSS J1027+1749 N & $-$0.180$\pm$0.007 & $-$0.456$\pm$0.013 & $-$0.283$\pm$0.025 & Composite & AGN & \cite{Liu2011b}\\ 
	SDSS J1027+1749 W & $-$0.008$\pm$0.009 & $-$0.215$\pm$0.004& 0.079$\pm$0.036 & LINER & N.D. & \cite{Liu2011b}\\ 
	\hline 
	NGC 3341 & 0.107$\pm$0.007 & $-$0.180$\pm$0.013 & 1.156$\pm$0.003 & AGN & AGN & \cite{Barth2008} \\ 
	NGC 3341 SW & $-$0.267$\pm$0.011 & $-$0.509$\pm$0.014& $-$0.420$\pm$0.003 & Composite & N.D. & \cite{Barth2008}\\ 
	NGC 3341 NW & $-$0.031$\pm$0.126 & $<-$0.221 & \dots  & \dots & N.D. & \cite{Barth2008} \\ 
	\hline 
	SDSS J1631+2352 & 0.348$\pm$0.002 & $-$0.019$\pm$0.003 & 0.891$\pm$0.004 & AGN & AGN &  This work \\
	\hline 
	SDSS J1708+2153 & $-$0.501$\pm$0.033 & $-$1.305$\pm$0.704 & 0.539$\pm$0.004 & AGN & AGN & This work  \\ 
	SDSS J1708+2153  NE & $-$0.148$\pm$0.035 & $-$0.364$\pm$0.073 & $-$0.41$\pm$0.40 & Composite & AGN & This work \\ 
	\hline
	SDSS J2356$-$1016 & $-$0.533$\pm$0.013 & $-$1.071$\pm$0.055 & 0.526$\pm$0.001 & AGN & AGN & This work \\ 
	SDSS J2356$-$1016 SE & $-$0.405$\pm$0.011 & $-$0.667$\pm$0.021 & 0.270$\pm$0.003 & Composite & AGN &  This work\\ 
	\hline 
	\hline
\end{tabular*}
\end{center}
Note. -- Columns: (1) Galaxy name; (2) $\log{}$[\ion{N}{2}]/H$\alpha$ ratio; (3) $\log{}$[\ion{O}{3}]/H$\beta$ ratio; (4) BPT classification, using the definitions presented in \cite{Kewley2001, Kauffmann2003, Ho1997b} and shown in Figure~\ref{fig:BPT}; (5) X-ray classification, as analyzed in this paper. AGN are defined as X-ray point sources with 2--7 keV luminosities greater than $10^{40}$ erg s$^{-1}$, with X-ray emission greater than expected from galactic XRBs; (6) References used for narrow emission-line ratios. N.D. denotes X-ray non-detections at a given galactic nucleus. NGC 3341 NW was found to have an unconstrained value of $\log{}$[\ion{O}{3}]/H$\beta$, due to the [\ion{O}{3}] $\lambda$5007 emission-line being undetected; furthermore, $\log{}$[\ion{O}{3}]/H$\beta$ On the top panel of Figure~\ref{fig:BPT}, we represent NGC 3341 NW with a dotted line. \\
\end{table*}
%
%
\par Of the 21 galactic nuclei (7 triple galaxy mergers), 15 have suitable spectroscopic coverage to analyze the BPT classifications. Four of the triple galaxy mergers (SDSS J0849+1114, SDSS J0858+1822, SDSS J1027+1749, and NGC 3341) received follow-up optical spectroscopic observations of each nucleus (see Table~\ref{tab:BPT} for more details), which we use in place of the SDSS DR16 measurements.

\par
\begin{figure}
\centering
    \includegraphics[width=0.5\textwidth]{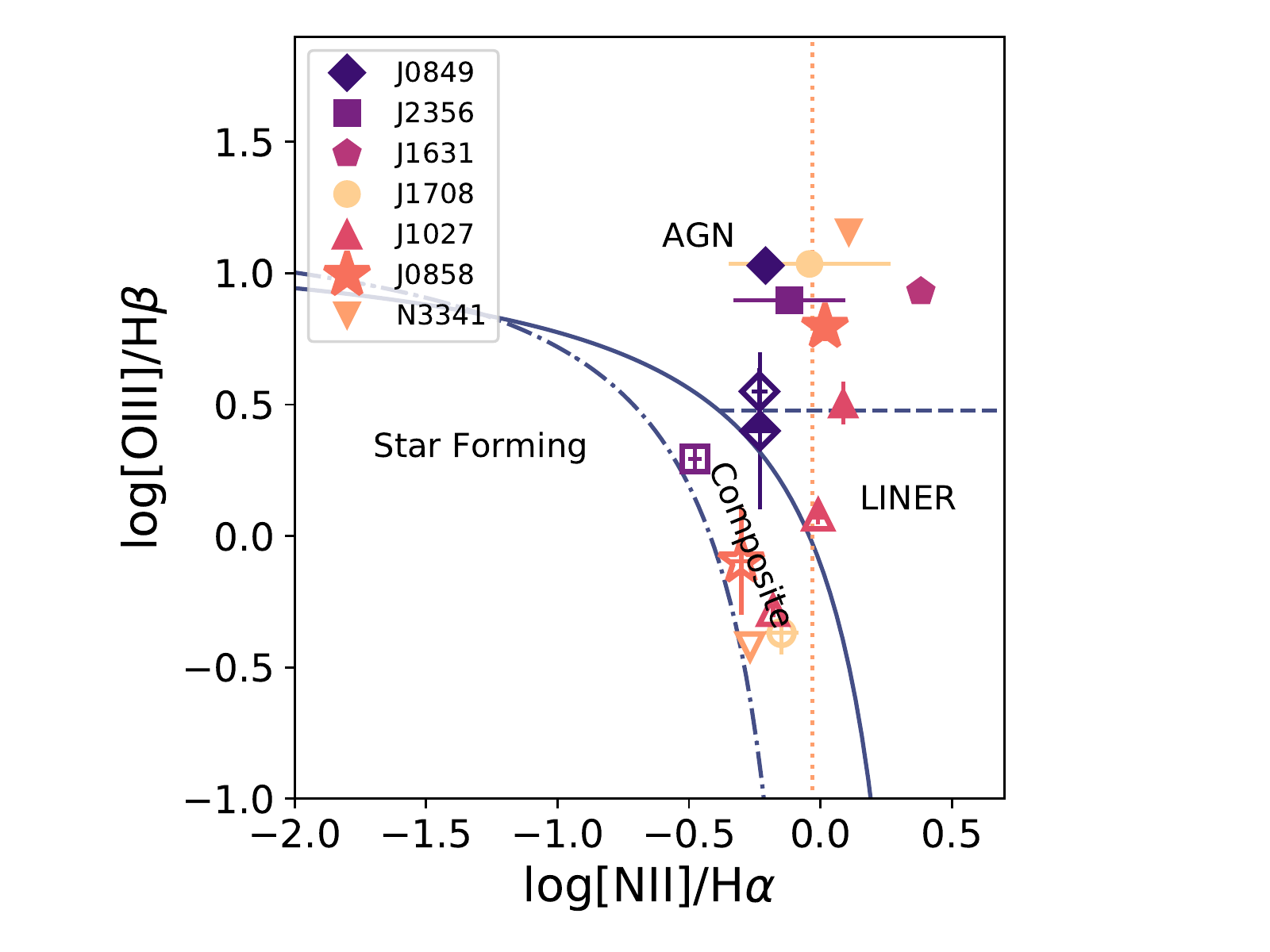}
    \includegraphics[width=0.5\textwidth]{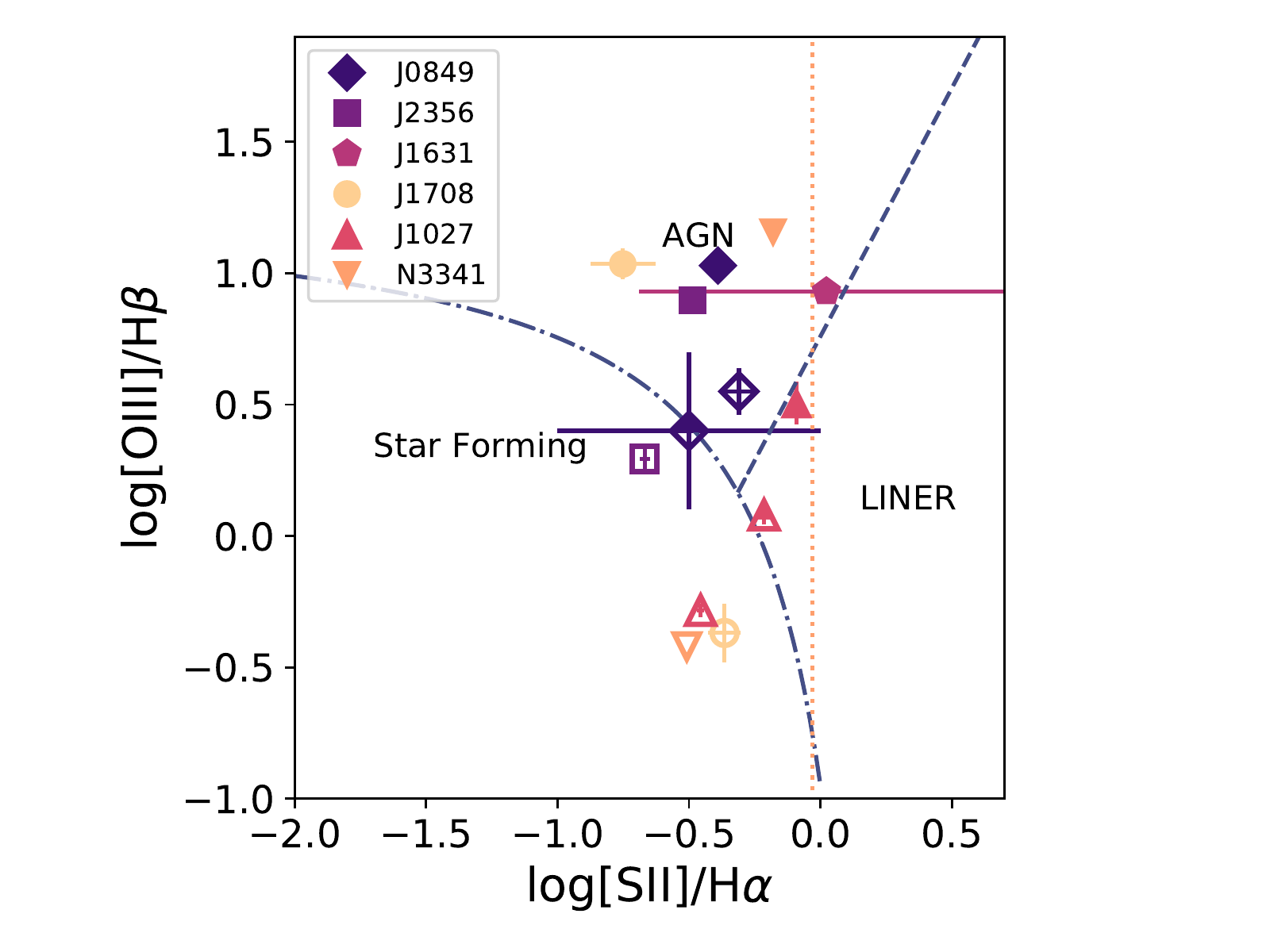}
    \vspace{-0.3cm}
    \caption{BPT optical spectroscopic line ratio diagrams for 15/21 galaxies in our triple merger sample, where blue lines represent the \cite{Kewley2001} (solid), \cite{Kauffmann2003} (dot-dashed), and \cite{Ho1997} (dashed) demarcations, which separate different sources of photoionization. For each triple merger system (differentiated by shape and color), we denote the primary (filled marker), secondary (unfilled marker), and tertiary (top-filled marker) X-ray point. We find that all primary X-ray AGN have nuclei classified as ``AGN" via BPT diagnostics, likely an effect of our sample selection technique, where 4/7 triple mergers were initially identified via the AllWISE AGN Catalog. Due to the [\ion{O}{3}] $\lambda$5007 emission-line being undetected, NGC 3341 NW is represented with a dotted line. We find instances where X-ray AGN are classified as either LINERs or Composite regions via BPT diagnostics, as well as non X-ray AGN classified as AGN via BPT diagnostics. These mismatches in AGN diagnostics can be due to both optical and X-ray emission being affected by obscuration, and/or shocks associated with the mergers.}

\label{fig:BPT}
\end{figure}

 We fit the SDSS spectra following a procedure comparable to the one presented in \cite{Tremonti2004}. We corrected the spectra for Galactic extinction using the \cite{Schlegel1998} dust maps and \cite{Cardelli1989} dust law. We shifted the spectra to rest-frame wavelengths, and identified the best-fitting model using chi-squared minimization. The model parameters were a Bruzual and Charlot starburst template (e.g., \citealt{Cales2013}) and Gaussians for the narrow emission-lines. The spectral fits for SDSS J1708+2153 and SDSS J2356$-$1016 also include an Fe II template from \cite{Boroson1992}. For the narrow-lines, we required that all Gaussians have the same width, that forbidden doublets have fixed wavelength and flux ratios, and the Balmer lines have fixed wavelength ratio. Broad-lines are fit with up to two Gaussians, with no constraints on their properties, and the AGN continuum is fit with a power law. Following the criteria presented in \cite{Hao2005} (see also \citealt{Reines2013, Baldassare2016}), we classify the galaxy as having a statistically significant broad emission-line if (1) $\chi_{\nu}$ is improved by at least 20\% when including the component and (2) the best-fit FWHM for the component is $>$ 500 km s$^{-1}$. We find statistically significant broad emission-line components (H$\alpha$ and/or  H$\beta$) in the spectra of SDSS J1631+2352, SDSS J1708+2153, SDSS J2356$-$1016, and SDSS J2356$-$1016 SE. Uncertainties on the Gaussian fluxes were obtained directly from the error matrix \citep{Kriss1994}. For SDSS J1708+2153 NE, we find that the H$\beta$ and [\ion{O}{3}] narrow emission-lines are weak, and thus the [\ion{O}{3}]/H$\beta$ measurements are likely less certain than the formal fit uncertainties. In Tables~\ref{tab:BPT} we list the narrow-line flux ratios for each nucleus with available data, while in Figure~\ref{fig:BPT} we plot the optical narrow line ratios. We find that all primary X-ray AGN detected by \BAYMAX{} are classified as AGN via BPT diagnostics.
\par \cite{Husemann2020} recently demonstrated that BPT diagnostics based on fiber spectroscopy from SDSS may lead to misclassifications when the spatial separation between sources is below 3\arcsec. At such separations, the spillover of flux from the dominate source into the secondary's SDSS fiber can be more than an order of magnitude. They found that the majority of the dual AGN candidates in their sample had secondary galaxies that were incorrectly classified as AGN when using SDSS measurements, compared to their results via follow-up long slit observations with the LBT (including SDSS J0858+1822). Of the three mergers that had their SDSS spectra analyzed, two (SDSS J1708+2153 and SDSS J2356$-$1016) have measurements for the secondary galaxy. However, we note that the primary and secondary galaxy in these mergers have angular separations greater than 3\arcsec~(6.6\arcsec and 3.5\arcsec, respectively), and they are both classified as Composite regions via BPT diagnostics. Given that SDSS J2356$-$1016 SE is only 3.5\arcsec~away from the primary galaxy, it is possible that the narrow emission-line ratios measured are even lower than the values presented in Table~\ref{tab:BPT}; this could possibly push SDSS J2356$-$1016 SE into the Star Forming regime of the BPT diagram.
%
%
\par One of the triple galaxy mergers in our sample, SDSS J1027+1749, was classified as a triple AGN system in \cite{Liu2011b} based on estimates of the $2-10$ keV luminosities (using the $L_{\mathrm{2-7 \ keV}}/L_{\mathrm{[OIII]}}$ ratios calculated for obscured AGN in \citealt{Panessa2007}). Analyzing the X-ray observation of the system with \BAYMAX{}, however, we classified this triple merger as a dual AGN candidate (see Section~\ref{chap6:origin}). Interestingly, the one nucleus with no X-ray point source is classified as a LINER via the BPT diagnostics shown in Figure~\ref{fig:BPT}, using ratios [\ion{N}{2}] $\lambda$6584/H$\alpha$, [\ion{S}{2}] $\lambda$6717,6731/H$\alpha$ or [\ion{O}{1}] $\lambda$6300/H$\alpha$ versus [\ion{O}{3}] $\lambda$5007/H$\beta$. The power source that is ionizing the gas in LINERs remains under debate, and has been explained in the past by low-luminosity active galactic nuclei (LLAGNs; \citealt{Ferland1983}), photoionization associated with stellar activity (e.g., \citealt{Barth2000}), and/or collisional or photo-excitation associated with shocks (e.g., \citealt{Heckman1980}). Although the majority of LINERs are believed to be associated with AGN (e.g., \citealt{Ho1997b}), these classifications have been based off complementary radio and X-ray detections. According to \cite{Molina2018}, the characteristic LINER spectrum can be created by a combination of these different physical mechanisms that get mixed in the spectroscopic fiber/aperture. The fact that there is no X-ray emission associated with a point source at the location of the nucleus lends credence to the theory that the photoionization in this nucleus is due to stellar or shock heating. On the other hand, AGN in galaxy mergers have been measured to have lower $L_{\mathrm{2-7 \ keV}}/L_{\mathrm{[OIII]}}$ ratios than non-merging AGN (see, e.g., \citealt{Liu2013, Comerford2015, Gross2019, Hou2020}), as well as weak correlations between the extinction corrected [\ion{O}{3}] and hard X-ray luminosities \citep{Berney2015,Lamastra2009,Marinucci2012}. These results are thought to be due to higher levels of obscuration at the smallest scales, geometrical differences in the scattering of the ionized gas, and/or long-term AGN variability (see, e.g., \citealt{Urrutia2008, Koss2010, Treister2012, Schawinski2012, Ellison2013, Villforth2014, Satyapal2014, Glikman2015, Fan2016, Weston2017, Goulding2017, Barrows2018, Koss2018}). Whether or not the nucleus of this galaxy hosts a central AGN can be better understood with higher-energy X-ray observations, due to photons above 10 keV being less susceptible to high levels of obscuration than the 0.5$-$8 keV population observed by \emph{Chandra}. Unfortunately, the resolution of \emph{NuSTAR} (with an average half-power diameter of $\sim$58\arcsec) will not allow for a resolved observation of three galactic nuclei.
\par Another mismatch between the X-ray and optical diagnostics occurs for SDSS J0858+1822. Our analysis with \BAYMAX{} finds the X-ray emission consistent with no AGN (although, given the estimated X-ray emission expected from star formation, there is evidence for high levels of obscuration, see Section~\ref{chap6:subsecJ0858}), while the BPT analysis classifies the primary nucleus as an AGN. This misclassification can be a result of many scenarios. For example, the AGN classification of the primary may be more complicated, given that the emission-line peak is offset by 1\arcsec~from the position of the galactic nucleus. The original SDSS observations were likely biased as a result, given that a high fraction of the flux from the brightest emission-line source was likely captured by both SDSS fibers. Although the follow-up long-slit observations carried out by \cite{Husemann2020} aimed to reduce the effects of spillover flux, future observations via high-resolution optical IFUs may allow for better insight on the ionization mechanism behind the emission-lines of each nuclei. If this misclassification is not a result of spillover contamination, emission from shocked gas is also known to be able to produce [\ion{O}{3}]/H$\beta$ and [\ion{N}{2}]/H$\alpha$ ratios consistent with values along the mixing sequence, typically towards the AGN region of the BPT diagram \citep{Rich2010,Rich2011, Kewley2013}. This type of contamination can be possibly enhanced in merging systems, where shocks are prevalent \citep{Rich2011}. \cite{DAgostino2019} present a new diagnostic capable of separating emission from star formation, shocks, and AGN simultaneously using IFU data. Folding in the maximum velocity dispersion in each spaxel, a three-dimensional diagnostic diagram can be used to better interpret whether emission in the AGN region is due to shocks.
\subsubsection{AGN components in the SDSS spectrum of SDSS J2356$-$1016 SE}
\par Although the narrow emission-line ratios place SDSS J2356$-$1016 SE in the Composite region of the BPT diagram, we find conclusive evidence for a broad H$\alpha$ component and strong evidence for an AGN power law continuum component ($f_{\lambda} \propto \lambda^{-\alpha}$) in the spectrum. In particular, we isolate the spectrum of SDSS J2356$-$1016 SE between rest-frame 6200 \AA~and 7000 \AA, and compare the spectral fits between two models with and without a broad emission-line feature. We find that the addition of a broad emission-line component best-fit at rest-frame 6563 \AA, is statistically significant, improving $\chi_{v}$ by 40\%. Assuming that the presence of a broad emission-line is a signature of gas rotating around a nuclear SMBH, this detection supports the conclusions reached by the X-ray analysis. We also find that the addition of an AGN power law continuum component significantly improves the fit. We fit the continuum of SDSS J2356$-$1016 SE between rest-frame 3500~\AA~ and 4700~\AA, and compare the spectral fits between two models with and without a power law component. Generally, a power law component can well-represent an intrinsic optical disk continuum, and the presence of a power law may be a sign of an underlying AGN. We find that the addition of the power law component improves the $\chi_{v}$ by 38\%. We test if adding an older stellar population to the spectral model, in place of a power law component, can account for the higher-levels of continuum in the blue wavebands. However, this spectral fit results in the largest $\chi_{v}$ value between all three models. 

Given the separation from the primary galaxy, it is also possible that these detections are an effect of spillover flux. Using the SDSS PSF models provided by \cite{Husemann2020}, we estimate that the level spillover flux from the primary can be anywhere from 4\%$-$10\% (taking into account the 2\arcsec~seeing during the SDSS observation). Given that the flux ratio of the broad H$\alpha$ emission-line between SDSS J2356$-$1016 SE and SDSS J2356$-$1016 is approximately 5\%, we can not statistically separate the emission from spillover flux at a high confidence level. If the spectrum of SDSS J2356$-$1016 SE is indeed affected by spillover flux, we may expect to also detect a H$\beta$ broad emission-line component, given the detection in the primary galaxy (see Table~\ref{tab:Broad}). Assuming a contamination fraction of $\approx$5\%, we find that a putative broad H$\beta$ emission-line in the spectrum of SDSS J2356$-$1016 SE would be undetected, given the levels of the stellar emission component.

\par
\begin{figure*}
\centering
    \includegraphics[width=12 cm]{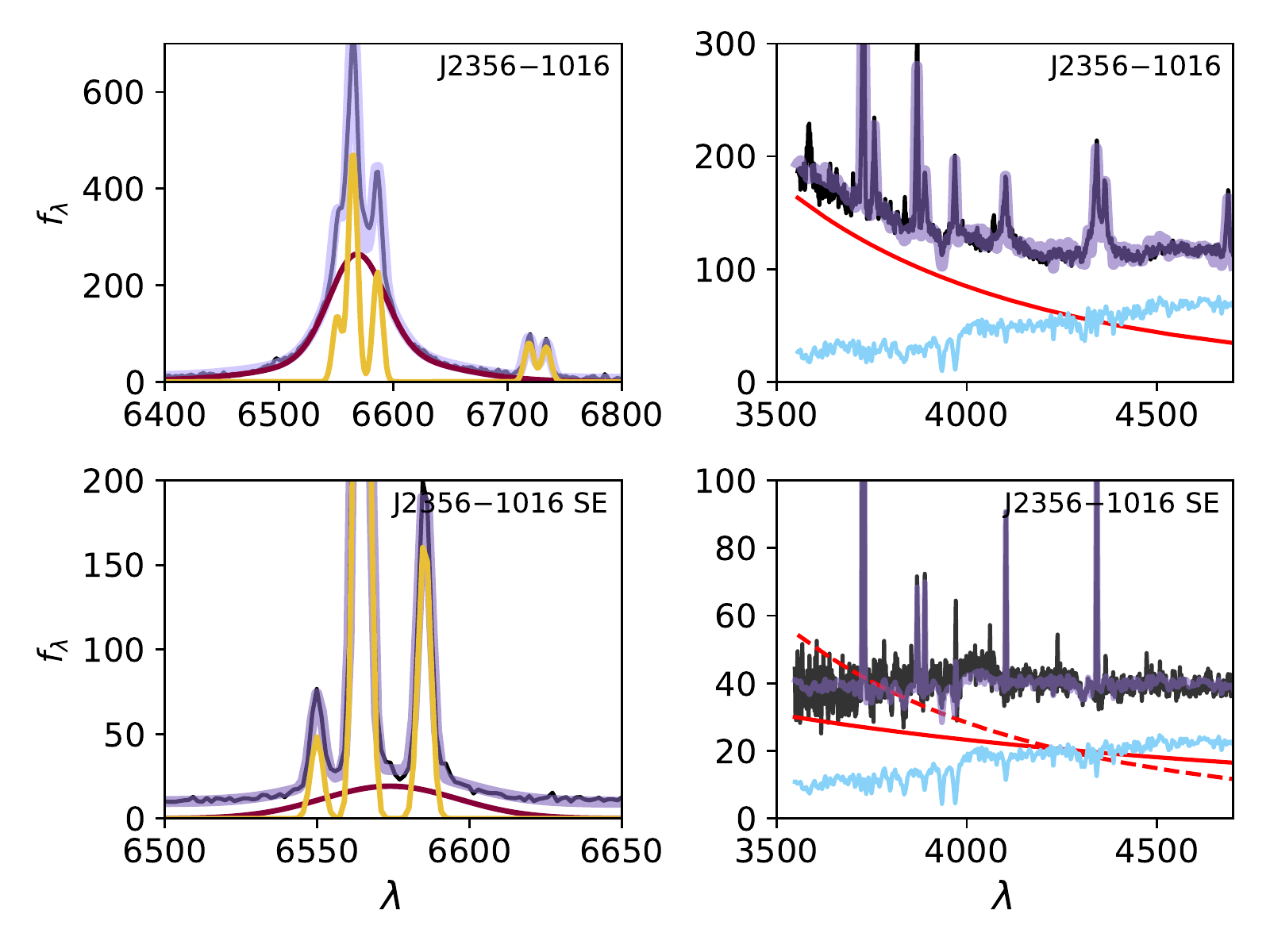}
    \vspace{-0.3cm}
    \caption{Flux density (in units of 10$^{-17}$ erg s$^{-1}$ cm$^{-2}$ \AA$^{-1}$) versus rest wavelength (in units of \AA) for SDSS J2356$-$1016 (top panels) and SDSS J2356$-$1016 SE (bottom panels). The spectra were fit using the procedure outlined in Section 4.2; the left panels show the observed emission (from which the stellar population model has been subtracted) between 6400 \AA~and 6800 \AA, while the right panels show the observed emission between 3500 \AA~and 4700 \AA. We decompose the spectra in the left panels into narrow emission-lines (orange) and broad emission-lines (dark red), while we decompose the spectra in the right panels into the stellar population component (light blue), and power law component (red). The combined emission model is shown in all panels in light purple. We plot a second power law component that has the same shape as that seen in the spectrum of SDSS J2356$-$1016, scaled down to match the levels seen in SDSS J2356$-$1016 (normalized at 4400 \AA, red dashed line). We find that the power law component of SDSS J2356$-$1016, assumed to originate from an underlying AGN, is significantly bluer than that of SDSS J2356$-$1016 SE, such that simply scaling the component of the brighter nucleus does not match the signature seen in SDSS J2356$-$1016. However, see Section 4.2.1 for a more detailed discussion on the possible origins of the power law spectral components for both SDSS J2356$-$1016 and SDSS J2356$-$1016 SE.}

\label{fig:Spectra}
\end{figure*}

 One major discrepancy between the AGN features seen in the spectra of SDSS J2356$-$1016 and SDSS J2356$-$1016 SE is the shape of their respective AGN power law continuum component. The power law continuum component of SDSS J2356$-$1016 (with best-fit spectral index $\alpha=5.58$) is significantly bluer than that of SDSS J2356$-$1016 SE (with best-fit spectral index $\alpha=2.12$), such that simply scaling down the component in the brighter nucleus does not match the signature seen in SDSS J2356$-$1016 SE (see Figure~\ref{fig:Spectra}). However, this discrepancy is a weak constraint, given that the power law component of SDSS J2356$-$1016 is extremely blue compared to the average AGN ($\alpha\sim0.5$, \citealt{Vandenberk2001}); this may be indicating that the power law component is representing multiple continuum components. Quasar spectra often rise strongly at $<$4000 \AA~due to the combination of UV Fe II emission and especially the Balmer continuum (e.g., \citealt{Wills1985}), which, if present, are all modeled as a single power law. However, if a Balmer continuum and Fe II components are causing the rise in the blue optical spectrum of SDSS J2356$-$1016, and the AGN spectral components measured in SDSS J2356$-$1016 SE are a result of spillover flux, than we expect these components are in the spectrum of SDSS J2356$-$1016 SE as well. As a result, the spectral index of the power law component would lowered for both spectra. Given the faint nature of the AGN power law continuum in J2356$-$1016 SE, constraining all these components to model the power-law component more faithfully is beyond the scope of these data.
\par Thus, although the AGN emission characteristics detected in SDSS J2356$-$1016 SE may be a result of spillover flux, the differences in the blue continuum of each galaxy tentatively suggests otherwise. Follow-up long-slit observations will be necessary to better understand possible AGN signatures in the spectrum of SDSS J2356$-$1016 SE. In Table~\ref{tab:Broad} we list the luminosities and FWHM of each detected broad emission-line component, and in Figure~\ref{fig:Spectra} we plot the SDSS spectral fits for both SDSS J2356$-$1016 and SDSS J2356$-$1016 SE.

%

\begin{table*}
\begin{center}
\caption{Broad Emission-Line Features}
\label{tab:Broad}
\begin{tabular*}{0.88\linewidth
}{lcccc}
	\hline
	\hline
	\multicolumn{1}{c}{Galaxy Name} &  \multicolumn{1}{c}{$\log{L_{\mathrm{H}\alpha}}$ (erg s$^{-1}$)} &
	\multicolumn{1}{c}{FWHM$_{\mathrm{H}\alpha}$ (km s$^{-1}$)} &
	\multicolumn{1}{c}{$\log{L_{\mathrm{H}\beta}}$ (erg s$^{-1}$)} &  \multicolumn{1}{c}{FWHM$_{\mathrm{H}\beta}$ (km s$^{-1}$)} \\
	\multicolumn{1}{c}{(1)} & \multicolumn{1}{c}{(2)} & \multicolumn{1}{c}{(3)} & \multicolumn{1}{c}{(4)} & \multicolumn{1}{c}{(5)} \\
	\hline
	SDSS J1631+2352 & 42.20 & 8760 & N/A & N.A  \\
	\hline 
	SDSS J1708+2153 & 42.10 & 5733 & 43.96 & 6768  \\ 
	\hline
	SDSS J2356$-$1016 & 42.03 & 2970 & 42.07 & 6769 \\
	SDSS J2356$-$1016 SE & 41.08 & 2401 & N/A & N/A  \\ 
	\hline
\end{tabular*}
\end{center}
Note. -- Columns: (1) Galaxy name; (2) H$\alpha$ luminosity of broad emission-line feature, in units of erg s$^{-1}$; (3) FWHM of best-fit H$\alpha$ broad emission-line feature, in units of km s$^{-1}$; (4) H$\beta$ luminosity of broad emission-line feature, in units of erg s$^{-1}$; (5) FWHM of best-fit H$\beta$ broad emission-line feature, in units of km s$^{-1}$.\\
\end{table*}

\section{Conclusions}
\label{chap6:conclusions}
In this study, we present a multi-wavelength analysis of AGN activity in triple galaxy mergers. We analyze 7 nearby ( $0.059 < z < 0.077$) triple galaxy mergers with existing archival \emph{Chandra}, SDSS DR16, \emph{WISE}, and VLA observations. Combining multiple wavelengths, we classify each detected X-ray point source with (i) $L_{\mathrm{2-7 \ keV}}>10^{40}$ erg s$^{-1}$ and (ii) $L_{\mathrm{2-7 \ keV}}$ greater than expected from the nuclear X-ray emission from XRBs as an AGN. Analyzing various parameters of each detected AGN, such as X-ray luminosity and levels of gas/dust obscuration, we further investigate differences in environments associated with single, dual, and triple AGN. The main results of this study are summarized below: 
\begin{enumerate}
    \item Combining the X-ray results from Foord et al. (2020b) with an analysis of the star formation rates of each galaxy, we find that all point sources have unabsorbed $2-7$ keV luminosities greater than 10$^{40}$ erg s$^{-1}$, and arguably greater than what is expected from nuclear (within 2\arcsec) contamination from HMXBs. The exception is the secondary point source in NGC 3341. Thus, we conclude that 1 triple merger systems in our sample is composed of a single AGN: NGC 3341; 4 triple merger systems in our sample are strong dual AGN candidates: SDSS J1027+1749, SDSS J1631+2352, SDSS J1708+2153, and SDSS J2356$-$1016; we confirm one triple AGN system, SDSS J0849+1114; and 1 triple merger in our sample remains ambiguous: SDSS J0858+1822.
    \item Analyzing the mid-IR emission of 6 of the mergers with archival \emph{WISE} observations, we find that only 4 are identified as AGN via standard mid-IR color-color AGN diagnostics. Furthermore, there does not appear to be a link between $W1-W2$ colors (representing how red the environment is) and total X-ray luminosity of the AGN in each system. However, accurate black holes mass measurements (and thus Eddington fraction measurements) of each system may give more insight on possible trends between the X-ray luminosities and $W1-W2$ colors.
    \item We find a trend of increasing $N_{H}$ (associated with the primary) as a function of increasing $W1-W2$ color. This is likely reflecting that the motions of gas and dust are coupled in merging environments, and large amounts of both can be funneled into the active central region during mergers. Interestingly, we find that the one triple AGN system in our sample has the highest levels of $N_{H}$ and $W1-W2$ colors, while the dual AGN all have lower levels. Given that SMBHs grow, and ignite to AGN, through the accretion of cool gas, it is possible that triple merger systems with higher levels of nuclear gas (measured by $N_{H}$) will have more AGN. Investigating whether or not this trend varies with total number of AGN in a given merger will require a larger sample of multiple AGN systems.
    \item We compare X-ray and optical AGN diagnostics by analyzing the placement of 15/21 galactic nuclei on the BPT diagram. We find that most primary X-ray AGN are also optically classified as AGN via BPT diagnostics, likely an effect of our sample selection technique. Mismatches include X-ray AGN being classified as Composite regions (SDSS J1027+1749, SDSS J1708+2153, and SDSS J2356$-$1016), as well as nuclei classified as AGN via BPT diagnostics but where \BAYMAX{} finds no X-ray emission consistent with a point source (SDSS J0858+1822). These mismatches in classification can be better understood with optical IFU observations.
\end{enumerate}

\par In general, given the larger ($>1\arcsec$) separations of these mergers, all multiple X-ray AGN candidates will benefit from further study using high quality optical spectroscopy via IFU. Tighter constraints on the [\ion{O}{3}] emission can be compared to typical AGN with X-ray emission (e.g. \citealt{Berney2015}) to confirm their dual AGN nature. Additionally, observations with ALMA or NIR IFU will allow for accurate measurements on the nuclear SFRs (which will not be as susceptible to obscuration as optical wavebands). Lastly, measurements of the black hole masses using velocity dispersions will provide helpful constraints on black hole growth, given the levels of detected X-ray emission. \\

\noindent \begin{center}{\MakeUppercase{\small Acknowledgements}}\end{center}
\vspace{-0.3cm}
\noindent We thank our referee for thorough and thoughtful feedback, which strengthened our analysis and results. A.F. and K.G. acknowledge support provided by the National Aeronautics and Space Administration through Chandra Award Numbers TM8-19007X, GO7-18087X, and GO8-19078X, issued by the Chandra X-ray Observatory Center, which is operated by the Smithsonian Astrophysical Observatory for and on behalf of the National Aeronautics Space Administration under contract NAS8-03060. A.F. also acknowledges support provided by the National Aeronautics and Space Administration through Chandra proposal ID 21700319. MK acknowledges support from NASA through ADAP award 80NSSC19K0749. The scientific results reported in this article are based on data obtained from the \emph{Chandra} Data Archive. This publication makes use of data produces from the Sloan Digital Sky Survey. Funding for the Sloan Digital Sky Survey has been provided by the Alfred P. Sloan Foundation, the U.S. Department of Energy Office of Science, and the Participating Institutions. This publication also makes use of data products from the Wide-field Infrared Survey Explorer, which is a joint project of the University of California, Los Angeles, and the Jet Propulsion Laboratory/California Institute of Technology, funded by the National Aeronautics and Space Administration. This research has
made use of NASA’s Astrophysics Data System.

\bibliographystyle{aasjournal}
\bibliography{foord.bib}

\end{document}